\documentclass[aps,prc,reprint,showpacs,floatfix,nofootinbib,superscriptaddress]{revtex4-1}
\usepackage{amsmath}
\usepackage{amssymb}
\usepackage{graphicx}
\usepackage{epsfig}

\newcommand{\fpi}{\mbox{$F_\pi$}}
\newcommand{\qsq}{\mbox{$Q^2$}}
\newcommand{\sigl}{\mbox{$\sigma_L$}}
\newcommand{\sigt}{\mbox{$\sigma_T$}}
\newcommand{\siglt}{\mbox{$\sigma_{LT}$}}
\newcommand{\sigtt}{\mbox{$\sigma_{TT}$}}

\newcommand{\gevsq}{\mbox{GeV$^2$}}

\graphicspath{{/home/huberg/r2d2/piminus/paper/}}
\DeclareGraphicsExtensions{ .eps , .eps.gz ,
                            .ps  , .ps.gz }

\begin{document}
\verb| |\\  

\title{Separated Response Functions in\\
Exclusive, Forward $\bf{\pi}^{\pm}$ Electroproduction on Deuterium}

\author{G.M.~Huber}
\affiliation{University of Regina, Regina, Saskatchewan S4S 0A2, Canada}
\author{H.P.~Blok}
\affiliation{VU university, NL-1081 HV Amsterdam, The Netherlands}
\affiliation{NIKHEF, Postbus 41882, NL-1009 DB Amsterdam, The Netherlands}
\author {C.~Butuceanu}
\affiliation{University of Regina, Regina, Saskatchewan S4S 0A2, Canada}
\author{D.~Gaskell}
\affiliation{Thomas Jefferson National Accelerator Facility,
 Newport News, Virginia 23606}
\author{T.~Horn}
\affiliation{Catholic University of America, Washington, DC 20064}
\author{D.J.~Mack}
\affiliation{Thomas Jefferson National Accelerator Facility,
 Newport News, Virginia 23606}
\author{D.~Abbott}
\affiliation{Thomas Jefferson National Accelerator Facility,
 Newport News, Virginia 23606}
\author{K.~Aniol}
\affiliation{California State University Los Angeles, Los Angeles, California
  90032}
\author{H.~Anklin}
\affiliation{Florida International University, Miami, Florida 33119}
\affiliation{Thomas Jefferson National Accelerator Facility,
 Newport News, Virginia 23606}
\author{C.~Armstrong}
\affiliation{College of William and Mary, Williamsburg, Virginia 23187}
\author{J.~Arrington}
\affiliation{Physics Division, Argonne National Laboratory, Argonne, Illinois
  60439}
\author{K.~Assamagan}
\affiliation{Hampton University, Hampton, Virginia 23668}
\author{S.~Avery}
\affiliation{Hampton University, Hampton, Virginia 23668}
\author{O.K.~Baker}
\affiliation{Hampton University, Hampton, Virginia 23668}
\affiliation{Thomas Jefferson National Accelerator Facility,
 Newport News, Virginia 23606}
\author{B.~Barrett}
\affiliation{Saint Mary's University, Halifax, Nova Scotia B3H 3C3 Canada}
\author{E.J.~Beise}
\affiliation{University of Maryland, College Park, Maryland 20742}
\author{C.~Bochna}
\affiliation{University of Illinois, Champaign, Illinois 61801}
\author{W.~Boeglin}
\affiliation{Florida International University, Miami, Florida 33119}
\author{E.J.~Brash}
\affiliation{University of Regina, Regina, Saskatchewan S4S 0A2, Canada}
\author{H.~Breuer}
\affiliation{University of Maryland, College Park, Maryland 20742}
\author{C.C.~Chang}
\affiliation{University of Maryland, College Park, Maryland 20742}
\author{N.~Chant}
\affiliation{University of Maryland, College Park, Maryland 20742}
\author{M.E.~Christy}
\affiliation{Hampton University, Hampton, Virginia 23668}
\author{J.~Dunne}
\affiliation{Thomas Jefferson National Accelerator Facility,
 Newport News, Virginia 23606}
\author{T.~Eden}
\affiliation{Thomas Jefferson National Accelerator Facility,
 Newport News, Virginia 23606}
\affiliation{Norfolk State University, Norfolk, Virginia 23504}
\author{R.~Ent}
\affiliation{Thomas Jefferson National Accelerator Facility,
 Newport News, Virginia 23606}
\author{H.~Fenker}
\affiliation{Thomas Jefferson National Accelerator Facility,
 Newport News, Virginia 23606}
\author{E.F.~Gibson}
\affiliation{California State University, Sacramento, California 95819}
\author{R.~Gilman}
\affiliation{Rutgers, The State University of New Jersey, Piscataway, New 
Jersey 08854}
\affiliation{Thomas Jefferson National Accelerator Facility,
 Newport News, Virginia 23606}
\author{K.~Gustafsson}
\affiliation{University of Maryland, College Park, Maryland 20742}
\author{W.~Hinton}
\affiliation{Hampton University, Hampton, Virginia 23668}
\author{R.J.~Holt}
\affiliation{Physics Division, Argonne National Laboratory, Argonne, Illinois
  60439}
\author{H.~Jackson}
\affiliation{Physics Division, Argonne National Laboratory, Argonne, Illinois
  60439}
\author{S.~Jin}
\affiliation{Kyungpook National University, Daegu, 702-701, Republic of Korea}
\author{M.K.~Jones}
\affiliation{College of William and Mary, Williamsburg, Virginia 23187}
\author{C.E.~Keppel}
\affiliation{Hampton University, Hampton, Virginia 23668}
\affiliation{Thomas Jefferson National Accelerator Facility,
 Newport News, Virginia 23606}
\author{P.H.~Kim}
\affiliation{Kyungpook National University, Daegu, 702-701, Republic of Korea}
\author{W.~Kim}
\affiliation{Kyungpook National University, Daegu, 702-701, Republic of Korea}
\author{P.M.~King}
\affiliation{University of Maryland, College Park, Maryland 20742}
\author{A.~Klein}
\affiliation{Old Dominion University, Norfolk, Virginia 23529}
\author{D.~Koltenuk}
\affiliation{University of Pennsylvania, Philadelphia, Pennsylvania 19104}
\author{V.~Kovaltchouk}
\affiliation{University of Regina, Regina, Saskatchewan S4S 0A2, Canada}
\author{M.~Liang}
\affiliation{Thomas Jefferson National Accelerator Facility,
 Newport News, Virginia 23606}
\author{J.~Liu}
\affiliation{University of Maryland, College Park, Maryland 20742}
\author{G.J.~Lolos}
\affiliation{University of Regina, Regina, Saskatchewan S4S 0A2, Canada}
\author{A.~Lung}
\affiliation{Thomas Jefferson National Accelerator Facility,
 Newport News, Virginia 23606}
\author{D.J.~Margaziotis}
\affiliation{California State University Los Angeles, Los Angeles, California
  90032}
\author{P.~Markowitz}
\affiliation{Florida International University, Miami, Florida 33119}
\author{A.~Matsumura}
\affiliation{Tohoku University, Sendai, Japan}
\author{D.~McKee}
\affiliation{New Mexico State University, Las Cruces, New Mexico 88003-8001}
\author{D.~Meekins}
\affiliation{Thomas Jefferson National Accelerator Facility,
 Newport News, Virginia 23606}
\author{J.~Mitchell}
\affiliation{Thomas Jefferson National Accelerator Facility,
 Newport News, Virginia 23606}
\author{T.~Miyoshi}
\affiliation{Tohoku University, Sendai, Japan}
\author{H.~Mkrtchyan}
\affiliation{A.I. Alikhanyan National Science Laboratory, Yerevan 0036,
  Armenia}
\author{B.~Mueller}
\affiliation{Physics Division, Argonne National Laboratory, Argonne, Illinois
  60439}
\author{G.~Niculescu}
\affiliation{James Madison University, Harrisonburg, Virginia 22807}
\author{I.~Niculescu}
\affiliation{James Madison University, Harrisonburg, Virginia 22807}
\author{Y.~Okayasu}
\affiliation{Tohoku University, Sendai, Japan}
\author{L.~Pentchev}
\affiliation{College of William and Mary, Williamsburg, Virginia 23187}
\author{C.~Perdrisat}
\affiliation{College of William and Mary, Williamsburg, Virginia 23187}
\author{D.~Pitz}
\affiliation{DAPNIA/SPhN, CEA/Saclay, F-91191 Gif-sur-Yvette, France}
\author{D.~Potterveld}
\affiliation{Physics Division, Argonne National Laboratory, Argonne, Illinois
  60439}
\author{V.~Punjabi}
\affiliation{Norfolk State University, Norfolk, Virginia 23504}
\author{L.M.~Qin}
\affiliation{Old Dominion University, Norfolk, Virginia 23529}
\author{P.E.~Reimer}
\affiliation{Physics Division, Argonne National Laboratory, Argonne, Illinois
  60439}
\author{J.~Reinhold}
\affiliation{Florida International University, Miami, Florida 33119}
\author{J.~Roche}
\affiliation{Thomas Jefferson National Accelerator Facility,
 Newport News, Virginia 23606}
\author{P.G.~Roos}
\affiliation{University of Maryland, College Park, Maryland 20742}
\author{A.~Sarty}
\affiliation{Saint Mary's University, Halifax, Nova Scotia B3H 3C3 Canada}
\author{I.K.~Shin}
\affiliation{Kyungpook National University, Daegu, 702-701, Republic of Korea}
\author{G.R.~Smith}
\affiliation{Thomas Jefferson National Accelerator Facility,
 Newport News, Virginia 23606}
\author{S.~Stepanyan}
\affiliation{A.I. Alikhanyan National Science Laboratory, Yerevan 0036,
  Armenia}
\author{L.G.~Tang}
\affiliation{Hampton University, Hampton, Virginia 23668}
\affiliation{Thomas Jefferson National Accelerator Facility,
 Newport News, Virginia 23606}
\author{V.~Tadevosyan}
\affiliation{A.I. Alikhanyan National Science Laboratory, Yerevan 0036,
  Armenia}
\author{V.~Tvaskis}
\affiliation{VU university, NL-1081 HV Amsterdam, The Netherlands}
\affiliation{NIKHEF, Postbus 41882, NL-1009 DB Amsterdam, The Netherlands}
\author{R.L.J.~van~der~Meer}
\affiliation{University of Regina, Regina, Saskatchewan S4S 0A2, Canada}
\author{K.~Vansyoc}
\affiliation{Old Dominion University, Norfolk, Virginia 23529}
\author{D.~Van~Westrum}
\affiliation{University of Colorado, Boulder, Colorado 80309}
\author{S.~Vidakovic}
\affiliation{University of Regina, Regina, Saskatchewan S4S 0A2, Canada}
\author{J.~Volmer}
\affiliation{VU university, NL-1081 HV Amsterdam, The Netherlands}
\affiliation{DESY, Hamburg, Germany}
\author{W.~Vulcan}
\affiliation{Thomas Jefferson National Accelerator Facility,
 Newport News, Virginia 23606}
\author{G.~Warren}
\affiliation{Thomas Jefferson National Accelerator Facility,
 Newport News, Virginia 23606}
\author{S.A.~Wood}
\affiliation{Thomas Jefferson National Accelerator Facility,
 Newport News, Virginia 23606}
\author{C.~Xu}
\affiliation{University of Regina, Regina, Saskatchewan S4S 0A2, Canada}
\author{C.~Yan}
\affiliation{Thomas Jefferson National Accelerator Facility,
 Newport News, Virginia 23606}
\author{W.-X.~Zhao}
\affiliation{Massachusetts Institute of Technology, Cambridge, 
Massachusetts 02139}
\author{X.~Zheng}
\affiliation{Physics Division, Argonne National Laboratory, Argonne, Illinois
  60439}
\author{B.~Zihlmann}
\affiliation{Thomas Jefferson National Accelerator Facility,
 Newport News, Virginia 23606}
\affiliation{University of Virginia, Charlottesville, Virginia 22901}
\collaboration{The Jefferson Lab \fpi\ Collaboration}
\noaffiliation

\date{\today}

\begin{abstract}
\begin{description}
\item[Background]
Measurements of forward exclusive meson production at different squared
four-momenta of the exchanged virtual photon, \qsq, and at different
four-momentum transfer, $t$, can be used to probe QCD's transition from
meson-nucleon degrees of freedom at long distances to quark-gluon degrees of
freedom at short scales.  Ratios of separated response functions in $\pi^-$ and
$\pi^+$ electroproduction are particularly informative.  The ratio for
transverse photons may allow this transition to be more easily observed, while
the ratio for longitudinal photons provides a crucial verification of the
assumed pole dominance, needed for reliable extraction of the pion form factor
from electroproduction data.
\item[Purpose]
Perform the first complete separation of the four unpolarized electromagnetic
structure functions $L$/$T$/$LT$/$TT$ in forward, exclusive $\pi^{\pm}$
electroproduction on deuterium above the dominant resonances.
\item[Method]
Data  were   acquired  with  2.6-5.2   GeV  electron  beams  and   the  HMS+SOS
spectrometers in  Jefferson Lab Hall  C, at central  \qsq\ values of  0.6, 1.0,
1.6~\gevsq\ at $W$=1.95 GeV, and $Q^2=2.45$ \gevsq\ at $W$=2.22 GeV.  There was
significant coverage in $\phi$ and $\epsilon$, which allowed separation of
$\sigma_{L,T,LT,TT}$.
\item[Results]
\sigl\ shows a clear signature of the pion pole, with a sharp rise at small
$-t$.  In contrast, \sigt\ is much flatter versus $t$.  The
longitudinal/transverse ratios evolve with \qsq\ and $t$, and at the highest
$Q^2=2.45$ \gevsq\ show a slight enhancement for $\pi^-$ production compared to
$\pi^+$.  The $\pi^-/\pi^+$ ratio for transverse photons exhibits only a small
\qsq-dependence, following a nearly universal curve with $t$, with a steep
transition to a value of about 0.25, consistent with $s$-channel quark
knockout.  The \sigtt/\sigt\ ratio also drops rapidly with \qsq, qualitatively
consistent with $s$-channel helicity conservation. The $\pi^-/\pi^+$ ratio for
longitudinal photons indicates a small isoscalar contamination at $W$=1.95 GeV,
consistent with what was observed in our earlier determination of the pion form
factor at these kinematics.
\item[Conclusions]
The separated cross sections are compared to a variety of theoretical
models, which generally describe \sigl\ but have varying success with \sigt.
Further theoretical input is required to provide a more profound insight into
the relevant reaction mechanisms for longitudinal and transverse photons, such
as whether the observed transverse ratio is indeed due to a 
transition from pion to quark knockout mechanisms, and provide useful
information regarding the twist-3 transversity generalized parton distribution, 
$H_T$.
\end{description}
\end{abstract}


\maketitle

\section{Introduction}

Exclusive electroproduction is a powerful tool for the study of
nucleon structure.  In contrast to inclusive $(e,e')$ or photoproduction
measurements, the transverse momentum of a scattering constituent (and thus its
transverse size is proportional to $1/\sqrt{-t}$) can be probed in addition to
its longitudinal momentum, and independent of the momentum transfer \qsq\ to
this constituent.  Exclusive {\em forward pion} electroproduction is especially
interesting because the longitudinal and transverse virtual photon
polarizations act as a filter on the spin and hence the type of the
participating constituents. By detecting the charge of the pion, even the
flavor of the constituents can be tagged.  Finally, {\em ratios} of separated
response functions can be formed for which nonperturbative corrections may
cancel, yielding insight into soft-hard factorization at the modest \qsq\ to
which exclusive measurements will be limited for the foreseeable future.  The
full potential of pion electroproduction is only now being realized due to the
availability of high-luminosity, multi-GeV beams at the Thomas Jefferson
National Accelerator Facility (Jefferson Lab, or JLab).

\begin{figure}
\begin{center}
\vskip 0.5cm
\includegraphics[angle=0,width=3.25in]{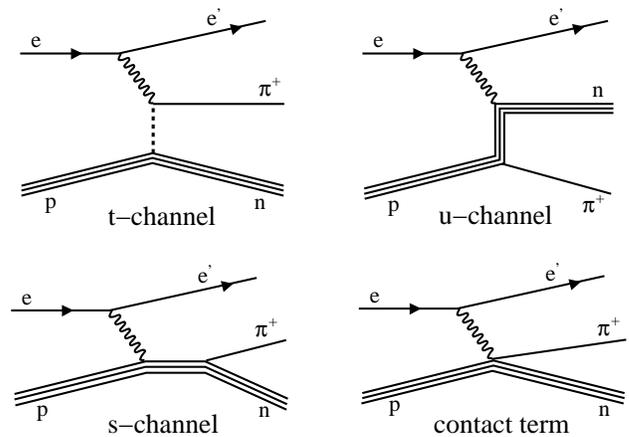}
\caption{\label{fig:stu_channels}
Born diagrams for $\pi^+$ electroproduction from a proton.}
\end{center}
\end{figure}

Four amplitudes contribute to pion electroproduction from a nucleon in the
Born approximation, where a single virtual photon $\gamma^*$
emitted by the electron couples to the hadronic system: pion-pole,
nucleon-pole, crossed nucleon-pole and contact term.  The first three
amplitudes correspond to Mandelstam $t$, $s$ and $u$-channel processes,
respectively, Fig. \ref{fig:stu_channels}.  The contact term is used to restore
gauge invariance.  Born-amplitude based models
~\cite{Ber70,Gut72} indicate that for values of the invariant mass $W$ above
the resonance region and for not too large values of \qsq, the longitudinal part
\sigl\ of the cross section for pion electroproduction at small values of
$-t$ is dominated by the $t$-channel process.  The other response functions
(transverse \sigt\ and interference terms \siglt\ and \sigtt)
are relatively small. In this regime, the process can be viewed
as quasi-elastic scattering of the electron from a virtual pion and thus is
sensitive to the pion form factor, $F_{\pi}$.  At values of $t$ approaching the
pion mass squared (the so called $t$-pole), the longitudinal response function
becomes approximately proportional to the square of the charged pion form
factor
\begin{equation} \sigma_{L} \approx
\frac{-t Q^2}{(t-M^2_{\pi})^2} g^2_{\pi NN}(t) F^2_{\pi}(Q^2, t).
\end{equation}  
Here, the factor $g_{\pi NN}(t)$ comes from the $\pi NN$ vertex and represents
the probability amplitude to have a virtual charged $\pi$ meson inside the
nucleon at a given $t$.

In order to reliably extract $F_{\pi}$, the $t$-pole process should be dominant
in the kinematic region under study.  This dominance can be studied
experimentally through the ratio of longitudinal $\gamma^{*}_L n \to \pi^- p$
and $\gamma^*_L p \to \pi^+ n$ cross sections, which can be expressed in terms
of contributions from isoscalar $A_S$ and isovector $A_V$ photon amplitudes:
\begin{equation}
R_L \equiv \frac{\gamma_L^{*}n \to \pi^- p}{\gamma_L^{*}p \to \pi^+ n} =
\frac{|A_{V}-A_{S}|^2}{|A_{V}+A_{S}|^2}.
\end{equation}
The $t$-channel process proceeds purely via isovector amplitudes.  Interference
terms between the isoscalar and isovector photon amplitudes have opposite signs
for $\pi^+$ and $\pi^-$ production, which leads to a difference in the cross
sections if significant isoscalar contributions are present.  Hence, where the
$t$-pole dominates (small $-t$), the ratio $R_L$ is expected to be close to
unity.  A departure from $R_L=1$ would indicate the presence of isoscalar
backgrounds arising from mechanisms such as $\rho$ meson exchange \cite{VGL1}
or perturbative contributions due to transverse quark momentum \cite{Milana}.
Such physics backgrounds may be expected to be larger at higher $-t$ (due to
the drop-off of the pion pole contribution) or non-forward kinematics (due to
angular momentum conservation) \cite{raskin}.  Because previous data are
unseparated \cite{Brauel1}, no firm conclusions about possible deviations of
$R_L$ from unity were possible.

One can also use such hard exclusive processes to investigate the range of
applicability of QCD factorization and scaling theorems.  The most important of
these is the handbag factorization, where only one parton participates in the
hard subprocess, and the soft physics is encoded in generalized parton
distributions (GPDs).  The handbag approach applies to deep exclusive meson
production, where \qsq\ is large and $-t$ is small \cite{collins,eides}.  For
longitudinal photons with $Q^2>10$ \gevsq\ and $-t\ll M_N^2$, this theorem allows
one to relate exclusive $N(e,e'\pi^{\pm})N$ observables to integrals over the
quark flavor-dependent GPDs.  Pseudoscalar
meson-production observables not dominated by the pion pole term, such as
\sigt\ in exclusive $\pi^{\pm}$ electroproduction, have also been identified as
being especially sensitive to the chiral-odd transverse GPDs \cite{ahmad,gk10}.
However, large higher-order corrections \cite{Belitsky} have delayed the
application of GPDs to pion electroproduction until recently.  The model of
Refs. \cite{gk10,gk13} uses a modified perturbative approach based on GPDs,
incorporating the empirical pion electromagnetic form factor and significant
contributions from the twist-3 transversity GPD, $H_T$, which gives substantial
strength in the transverse cross section.

In the transition region between low $-t$ (where a description of hadronic
degrees of freedom in terms of effective hadronic Lagrangians is valid) and
large $-t$ (where the degrees of freedom are quarks and gluons), $t$-channel
exchange of a few Regge trajectories permits an efficient description of the
energy-dependence and the forward angular distribution of many real- and
virtual-photon-induced reactions.  The VGL Regge model \cite{VGL,VGL1} has
provided a good and consistent description of a wide variety of $\pi^{\pm}$
photo- and electroproduction data above the resonance region, as well as of the
$p(e,e'\pi^+)n$ reaction using longitudinally polarized virtual photons.
However, the model has consistently failed to provide a good description of
$p(e,e'\pi^+)n$ \sigt\ data \cite{Blok08}.  The VGL Regge model was recently
extended \cite{kaskulov, vrancx} by the addition of a hard deep-inelastic
scattering (DIS) of virtual photons off nucleons.  The DIS process
dominates the transverse response at moderate and high \qsq, providing a better
description of \sigt.  By assuming that the exclusive \sigt\ cross section
behaves as $\sigma_T^{\text{DIS}}(Q^2)\propto F_1^p(x,Q^2)$, the authors
predict that at moderate \qsq
\begin{equation}
R_T\equiv\frac{\sigma_T^{\pi^-}}{\sigma_T^{\pi^+}}\approxeq\frac{F_1^n}{F_1^p}
\approx\frac{F_2^n}{F_2^p}<1.
\end{equation}

Our purpose was to perform a complete $L$/$T$/$LT$/$TT$ separation in exclusive
forward $\pi^{\pm}$ electroproduction on the proton and neutron.  Because there
are no practical free neutron targets, the $^2$H$(e,e'\pi^{\pm})NN_s$ reactions
(where $N_s$ denotes the spectator nucleon) were used.  As those reactions
proceed via quasi-free production, the results can be used to compare $\pi^-$
production on the neutron to $\pi^+$ production on the proton, particularly if
ratios are formed.  However, due to binding effects, the $\pi^+$ results on the
deuteron may differ from those on the proton, which were taken in the same
kinematics.  The data were obtained in Hall C at JLab as part of the two pion
form factor experiments presented in detail in Ref.~\cite{Blok08}.  The purpose
of this paper is to describe the experiment and analysis of these data in
detail, concentrating on those parts that differ from our $^1$H$(e,e'\pi^+)n$
study Ref.~\cite{Blok08}.

This paper is organized as follows.  Sec \ref{sec:expt} describes the
experiment and the determination of the various efficiencies that are applied
in calculating the cross sections.  Sec \ref{sec:analysis} presents the
determination of the unseparated cross sections, their separation into the
$L$/$T$/$LT$/$TT$ structure functions, and the systematic uncertainties.  Sec
\ref{sec:results} discusses these results and compares them with various
theoretical calculations.  The paper is concluded with a short summary.

\section{Experiment and Data Analysis
\label{sec:expt}}

The analysis procedures applied here were also used in our recent letter
on $^2$H$(e,e'\pi^{\pm})NN_s$ results
\cite{Huber14}.  For details of the experiment and the analysis not discussed
here, we refer the reader to the discussion of our $^1$H experiment
\cite{Blok08}.

\subsection{Experiment and Kinematics
\label{sec:kinematics}}

The two $F_{\pi}$ experiments were carried out in 1997 (F$_{\pi}$-1) and 2003
(F$_{\pi}$-2) in Hall C at JLab.  For the measurements presented here, the
unpolarized electron beam from the CEBAF accelerator was incident on a liquid
deuterium target. Two moderate acceptance, magnetic focusing spectrometers were
used to detect the particles of interest.  The spectrometer settings correspond
to either $^2$H$(e,e'\pi^+)nn_s$ or $^2$H$(e,e'\pi^-)pp_s$ kinematics, where the
Short Orbit Spectrometer (SOS) was always used to detect the scattered
electron, and the High Momentum Spectrometer (HMS) was used to detect the high
momentum $\pi^+$ or $\pi^-$.

The choice of kinematics for the two experiments was based on maximizing the
range in \qsq\ for a value of the invariant mass $W$ above the resonance
region, while still enabling a longitudinal-transverse separation.  The value
$W$=1.95 GeV used in the first experiment is high enough to suppress most
$s$-channel baryon resonance backgrounds, but this suppression should be even
more effective at the $W$=2.2 GeV of the second experiment.  For each \qsq,
data were taken at two values of the virtual photon polarization, $\epsilon$,
with $\Delta\epsilon>$0.25.  This allowed for a separation of the longitudinal
and transverse cross sections.  Constraints on the kinematics were imposed by
the maximum available electron energy, the maximum central momentum of the SOS,
and the minimum HMS angle.  In parallel kinematics, i.e., when the pion
spectrometer is situated in the direction of the $\vec{q}$ vector, the
acceptances of the two spectrometers do not provide a uniform coverage in
$\phi_{\pi}$.  Thus, to attain full coverage in $\phi_{\pi}$ and allow a
separation of the interference $LT$ and $TT$ cross sections, additional data
were taken in most cases with the HMS at a slightly smaller and larger angle
compared to the central angle for the high $\epsilon$ settings. At low
$\epsilon$, only the larger angle setting was possible.  The kinematic settings
are summarized in Table \ref{tab:kin}.

\begin{table*}
\begin{center}
\begin{tabular}{|c|c|c|}\hline
            & $^2$H$(e,e'\pi^+)nn$ & $^2$H$(e,e'\pi^-)pp$\\
\hline
\multicolumn{3}{|c|}{F$_{\pi}$-1 Settings}\\
\hline
\multicolumn{3}{|c|}{$Q^2$=0.6 GeV$^2$, $W=$1.95 GeV}\\
\hline
$\epsilon$=0.37, $E_{e}=$2.445 GeV & 3 HMS settings: $\Theta_{\pi q}$=+0.5,
+2.0, +4.0$^o$ & 2 HMS settings: +0.5, +4.0$^o$\\
$\epsilon$=0.74, $E_{e}=$3.548 GeV & 4 HMS settings: $\Theta_{\pi q}$=-2.7, 
0.0, +2.0, +4.0$^o$ & 1 HMS setting: 0.0$^o$\\
\hline
\multicolumn{3}{|c|}{$Q^2$=0.75 GeV$^2$, $W=$1.95 GeV}\\
\hline
$\epsilon$=0.43, $E_{e}=$2.673 GeV & 2 HMS settings: $\Theta_{\pi q}$=0.0,
+4.0$^o$ & 2 HMS settings: $\theta_{\pi q}$=0.0, +4.0$^o$\\
$\epsilon$=0.70, $E_{e}=$3.548 GeV & 3 HMS settings: $\Theta_{\pi q}$=-4.0, 
0.0, +4.0$^o$ & No data\\
\hline
\multicolumn{3}{|c|}{$Q^2$=1.0 GeV$^2$, $W=$1.95 GeV}\\
\hline
$\epsilon$=0.33, $E_{e}=$2.673 GeV & 2 HMS settings: $\Theta_{\pi q}$=0.0,
+4.0$^o$ & 2 HMS settings: $\theta_{\pi q}$=0.0, +4.0$^o$\\
$\epsilon$=0.65, $E_{e}=$3.548 GeV & 3 HMS settings: $\Theta_{\pi q}$=-4.0, 
0.0, +4.0$^o$ & 1 HMS setting: 0.0$^o$\\
\hline
\multicolumn{3}{|c|}{$Q^2$=1.6 GeV$^2$, $W=$1.95 GeV}\\
\hline
$\epsilon$=0.27, $E_{e}=$3.005 GeV & 2 HMS settings: $\Theta_{\pi q}$=0.0,
+4.0$^o$ & Same settings as $\pi^+$\\
$\epsilon$=0.63, $E_{e}=$4.045 GeV & 3 HMS settings: $\Theta_{\pi q}$=-4.0, 
0.0, +4.0$^o$ & Same settings as $\pi^+$\\
\hline
\multicolumn{3}{|c|}{F$_{\pi}$-2 Settings}\\
\hline
\multicolumn{3}{|c|}{$Q^2$=2.45 GeV$^2$, $W=$2.2 GeV}\\
\hline
$\epsilon$=0.27, $E_{e}=$4.210 GeV & 2 HMS settings: $\Theta_{\pi q}$=+1.35,
+3.0$^o$ & Same settings as $\pi^+$\\
$\epsilon$=0.55, $E_{e}=$5.248 GeV & 3 HMS settings: $\Theta_{\pi q}$=-3.0, 
0.0, +3.0$^o$ & Same settings as $\pi^+$\\
\hline
\end{tabular}
\caption{A summary of the $^2$H kinematic settings taken in the two
  pion form factor experiments.  The angle $\Theta_{\pi q}$ refers
  to the lab angle between the pion spectrometer and the central
  $\vec{q}$-vector as defined by the beam energy and the angle of the
  electron spectrometer.
\label{tab:kin}}
\end{center}
\end{table*}

For each \qsq, $\epsilon$ setting, the electron spectrometer angle and
momentum, as well as the pion spectrometer momentum, were kept fixed.  The HMS
magnetic polarity was reversed between $\pi^+$ and $\pi^-$ running, with the
quadrupoles and dipole magnets cycled according to a standard procedure, then
set to the final values by current (in the case of the quadrupoles) or by NMR
probe (in the case of the dipole).

Kinematic offsets in spectrometer angle and momentum, as well as in beam
energy, were previously determined using elastic $ep$ coincidence data taken
during the same run, and the reproducibility of the optics was checked
\cite{Blok08}.  For the deuterium data sets studied here, elastic runs on $^1$H
were used to check the validity of the HMS and SOS corrections for several
momentum ranges.  The reproducibility of the optics was checked during electron
running with sieve slits and by the position of the missing mass peak for
$^2$H$(e,e'\pi^+)nn_s$ or $^2$H$(e,e'\pi^-)pp_s$.  No shifts beyond the
expected calibration residuals $\pm$2 MeV were observed
\cite{Volmerthesis,Tanjathesis}.

\subsection{HMS Tracking and Tracking Efficiency
\label{sec:trackeff}}

The HMS singles rates were much higher for the $\pi^-$ settings than the
$\pi^+$ settings because of the large electron background at negative
spectrometer polarity, so accurate HMS track reconstruction at high rates is
needed.  Charged particle trajectories are measured by two drift chambers, each
with six planes \cite{chambers}.  All data presented here used the track
selection criterion that 5 out of 6 planes in each drift chamber for both
spectrometers should have a valid signal.  This criterion is much better
suited to high rate data (in this case the $\pi^-$ channel data) than the
analysis of our earlier F$_{\pi}$-1 $\pi^+$ data from hydrogen target
\cite{volmer,tadevosyan}, which used a 4/6 tracking selection criterion for HMS
and 5/6 for SOS tracking.

The HMS tracking algorithm used here is the same as used in our earlier
F$_{\pi}$-2 analysis from liquid hydrogen target \cite{hornt}.  The algorithm has
several requirements:
\begin{itemize}
\item {If the program reconstructed only one track, then that track was used.}
\item {If two or more tracks are reconstructed, then the track that
    projects to the blocks in the calorimeter measuring the energy deposit
(i.e. the cluster) was used.  The calorimeter cut used was quite
loose to only eliminate ``noise'' tracks in the chambers.}
\item{In case two or more tracks hit the cluster in the calorimeter (or neither
of them), then additional criteria based on which hodoscope bar was hit were
used to select a correct track.}
\end{itemize}
The above criteria ensured that the chosen track was the most likely one to
have resulted from the trigger for that event and greatly reduced the number of
events improperly excluded from the analysis.

The fiducial tracking inefficiencies were 2-9\% for HMS rates up to 1.4 MHz.
The tracking efficiency is defined as the ratio of the number of events for
which an actual track is found, to the `events' that pass through the drift
chambers.  This ratio is extracted from events in a fiducial area where it is
extremely likely that the scintillator hits are due to particles that also
traversed the chambers.  The tracking efficiency depends on both the drift
chamber hit efficiency and the tracking algorithm used in finding the track.

In order to accurately calculate the tracking efficiency, tight particle
identification (PID) requirements were applied to select a pure data
sample. These requirements are stricter than those used in the regular
analysis.  In the HMS, the particle identification requirements used to select
pions in the tracking efficiency calculation consisted of cuts on the gas
\u{C}erenkov and the calorimeter for F$_{\pi}$-1 data, while for F$_{\pi}$-2 an
additional cut on the aerogel \u{C}erenkov was applied.

The fiducial tracking efficiency analysis also incorporates a cut on the
integrated pulse (ADC) from the scintillator hodoscope PMTs, to exclude
events with multiple hits per scintillator plane.  In the case where there are
multiple tracks in the same scintillator plane, this cut places a bias on the
event sample used to calculate the tracking efficiency.  Since 2-track
events have a lower efficiency than 1-track events, the resulting bias causes
the HMS tracking efficiency to be overestimated.

To obtain a better understanding of the HMS tracking efficiencies, in
F$_{\pi}$-2 a study of singles yields from a carbon target versus HMS rate and
beam current was performed.  The
normalized yields from a carbon target should present no significant beam
current- or rate-dependence if the various efficiencies are calculated
correctly.  Unfortunately, no luminosity scans on carbon target were taken at
different beam currents in the F$_{\pi}$-1 experiment, so any conclusions
obtained from the F$_{\pi}$-2 study have to be applied also to the
F$_{\pi}$-1 data.

\begin{figure}
\begin{center}
\includegraphics[angle=89.9,width=3.25in]{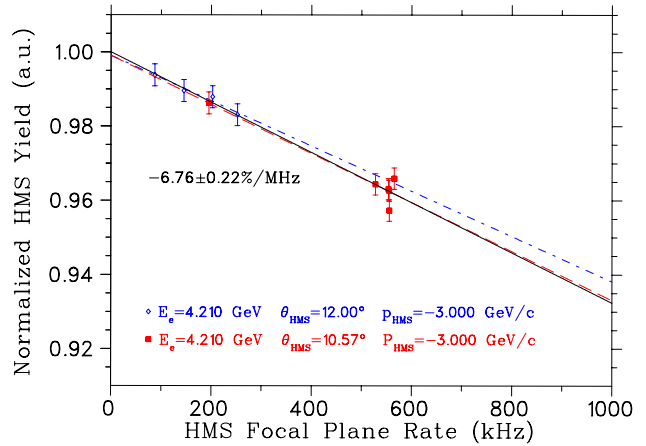}
\caption{\label{fig:Carbon_boiling_rate}
(Color online) Normalized yields (no PID cut) from the carbon target versus HMS
singles rate.  As the tracking efficiency calculation uses a data sample where
multiple track events are rejected, the HMS tracking efficiencies are
overestimated at high rates, leading to an effective drop in normalized yield
versus rate.  The HMS tracking efficiencies for both of the F$_{\pi}$-1 and
F$_{\pi}$-2 data sets are corrected with the linear rate dependent function
shown here, leading to a normalized yield that is independent of rate.
}
\end{center}
\end{figure}

Since the probability of a second particle traversing the HMS during the event
resolving time is greater at high rates, a tight electron PID cut might
introduce its own deadtime not due to tracking efficiency, causing the 
rate-dependence to be underestimated.  Therefore, only HMS fiducial acceptance
cuts were applied in this study.  Normalized yields from the carbon target were
computed from the number of events passing cuts, the integrated beam charge,
the electronic and CPU data acquisition livetimes and the HMS tracking
efficiency.  They are plotted versus rate in
Fig. \ref{fig:Carbon_boiling_rate}.  The error bars include statistical
uncertainties and an estimated systematic uncertainty of 0.3\% added in
quadrature, to take into account beam steering on the target and other
sensitive effects when no PID cut is applied.  Data from the two kinematic
settings were separately fit versus rate (dashed red and dash-dot blue curves
in the figure) and normalized to unity at zero rate.  The two data sets, thus
normalized, were then fit together, yielding the solid black curve.  The
observed rate-dependence suggests that the fiducial HMS tracking efficiencies
$htr$, as determined using the procedure described at the start of this
section, should be corrected in the following manner
\begin{equation}
htr^{\prime}=htr (e^{-6.76236\cdot 10^{-5}\times \text{HMSrate(kHz)}}).
\label{eqn:tracking_corr}
\end{equation}
This is particularly important for the F$_{\pi}$-1 $\pi^-$ runs, which are at
higher HMS rate.

The systematic uncertainties in the HMS tracking efficiencies were estimated as
follows.  In the F$_{\pi}$-2 hydrogen analysis, the tracking efficiencies were
assigned a 1.0\% scale and an 0.4\% $\epsilon$-uncorrelated systematic
uncertainty, where the first is the scale uncertainty common to all settings,
and the second is due to a variety of factors that may affect high and low
$\epsilon$ settings differently, as evidenced by the greater scatter exhibited
by the tracking efficiencies at high rates (see Refs. \cite{Tanjathesis,Blok08}
and Sec. \ref{sec:syst}).  There is an additional uncertainty of 0.2\%/MHz
due to the tracking efficiency correction shown in
Fig. \ref{fig:Carbon_boiling_rate}.  Since the maximum rate variation for all
F$_{\pi}$-2 $\pi^{\pm}$ settings, as well as the F$_{\pi}$-1 $\pi^+$ settings,
is about 400~kHz, this gives a total $\epsilon$-uncorrelated systematic
uncertainty of 0.45\%.  The F$_{\pi}$-1 $\pi^-$ $\epsilon$-uncorrelated
systematic uncertainty is somewhat larger.  Since the high rate scatter in
these $\pi^-$ tracking efficiencies is approximately $\pm 1.25\%$ at 1.3 MHz,
we assign an $\epsilon$-uncorrelated systematic uncertainty for these settings
of 1.3\%.

In addition to the above tracking efficiencies, the experimental yields were
also corrected for data acquisition electronic and CPU dead time.  The
correction ranged from 1-11\% with minimal uncertainty, as discussed in
Refs. \cite{Blok08,Volmerthesis}.

\subsection{Cryotarget Boiling Correction
\label{sec:ld2_boiling}
}

When the electron beam hits a liquid target, it deposits a large power
per unit target area and as a result induces localized density
fluctuations referred to as ``target boiling.''  In order to reduce these
fluctuations, the beam was rastered over a small area rather than localizing it
at one point on the target.  The target boiling effect can be measured by
comparing the yields at fixed kinematics and varying beam current.  During both
experiments (F$_{\pi}$-1 and F$_{\pi}$-2), dedicated luminosity elastic runs
were taken for both liquid targets (hydrogen and deuterium).  The two
experiments used cryotargets with significantly different geometries, as well
as significantly different beam raster patterns, leading to very different
boiling effects.

F$_{\pi}$-2 used the ``tuna can'' cryotarget geometry\footnote{Cylindrical
cryotarget with its axis vertical, transverse to the beam.}  and circular beam
raster design, which are expected to result in boiling corrections
$<1\%$~\cite{Tanjathesis}.  To determine the appropriate correction 
when the corrected HMS tracking efficiencies are used, data were
acquired in dedicated runs with a wide variety of electron beam currents for
all $\pi^-$ kinematic settings except \qsq=2.45~\gevsq, high $\epsilon$,
$E_e=5.25$ GeV, $\theta_{\text{HMS}}=13.61^{\circ}$.  Only fiducial acceptance
cuts were applied in this study, and normalized singles yields from these $^2$H
negative polarity HMS data were computed from the number of counts passing
cuts, the integrated beam charge, electronic and CPU data acquisition
livetimes, and the HMS tracking efficiencies corrected via
Eqn.~\ref{eqn:tracking_corr}.  The observed current-dependence suggests that no
correction should be applied, which is similar to the conclusion reached in
Ref.~\cite{Tanjathesis} for a liquid $^1$H target of the same shape and
dimensions.

\begin{figure}
\begin{center}
\includegraphics[angle=90,width=3.25in]{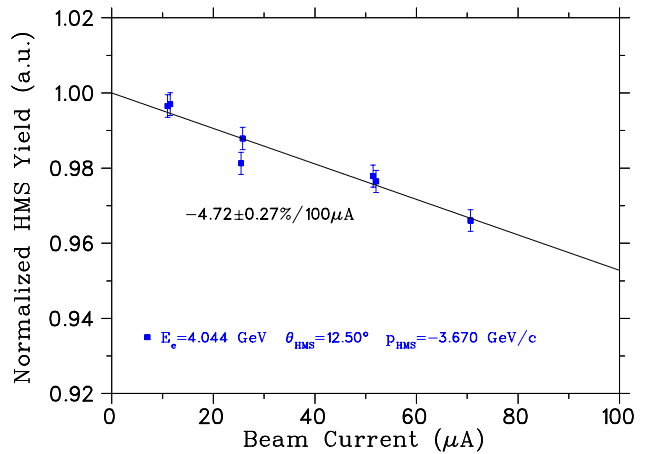}
\caption{\label{fig:Fpi1_LD2_boiling} 
(Color online) Normalized HMS yields from F$_{\pi}$-1 $^2$H elastics data taken
  with an electron trigger plotted as a function of beam current.  A +0.2$~\mu$A 
  beam current offset is applied, as described in the text.  The error bars
  include statistical uncertainties and an estimated systematic uncertainty of
  0.3\% added in quadrature.}
\end{center}
\end{figure}

F$_{\pi}$-1 used the so-called ``soda can'' cryotarget
geometry\footnote{Cylindrical cryotarget with its axis horizontal, in the
  direction of the beam.}  and ``bed post'' beam rastering\footnote{Un-even
  rastering over a rectangular area, with sinusoidal motion in $x$ and $y$,
  leading to the beam spending more time on the four corners and less time in
  the middle, see Fig~3.3 of Ref.~\cite{Volmerthesis}.}, which leads to a
significant boiling correction.  The magnitude of this correction is sensitive
to the rate-dependent correction applied to the HMS tracking efficiencies.  The
HMS tracking efficiencies were corrected via Eqn.~\ref{eqn:tracking_corr} and
normalized yields calculated in the same manner as in the F$_{\pi}$-2
cryotarget boiling study.  In analyzing these data, it was found that the slope
of yield versus beam current was overly sensitive to the inclusion of the
lowest current points in the fit.  The beam current calibration has an inherent
0.2~$\mu$A uncertainty due to noise in the Unser monitor.  A sigificantly
reduced sensitivity to these low current points was obtained with the addition
of a +0.2~$\mu$A beam current offset, which was subsequently applied in all
F$_{\pi}$-1 yield calculations was determined via a $\chi^2$ minimization
technique.  A similar current offset was used in Ref. ~\cite{Gaskellthesis}.

The corrected data were thus fit versus current and normalized to unity at zero
current, yielding the black curve in Fig. \ref{fig:Fpi1_LD2_boiling}, and a
$^2$H target density correction of $(4.72\pm 0.27\%)/100~\mu$A.  This
correction is particularly important for the F$_{\pi}$-1 $\pi^+$ data.  Since
the HMS detector rates were lower when the HMS was set at positive polarity
compared to negative polarity, higher incident electron beam currents were
often used for the $\pi^+$ runs.  The resulting cryotarget boiling correction
is similar to the $(6\pm 1\%)/100~\mu$A correction determined for the
F$_{\pi}$-1 $^1$H cell in Ref.~\cite{Volmerthesis}.

\subsection{HMS \u{C}erenkov Blocking Correction
\label{sec:cerblock}}

The potential contamination by electrons when the pion spectrometer is set to
negative polarity, and by protons when it is set to positive polarity,
introduces some differences in the $\pi^{\pm}$ data analyses which were
carefully examined.  For most negative HMS polarity runs, electrons were
rejected at the trigger level by a gas \u{C}erenkov detector containing
C$_4$F$_{10}$ at atmospheric pressure acting as a veto in order to avoid high
DAQ deadtime due to large $e^{-}$ rates in the HMS.  There is a loss of pions
due to electrons passing through the HMS gas \u{C}erenkov within $\sim$100~ns
after a pion has traversed the detector, resulting in a mis-identification of
the pion event as an electron and being eliminated by the PID cuts applied
(\u{C}erenkov blocking).  To reduce this effect, the beam current was
significantly reduced during $\pi^-$ running.  Two independent studies were
performed to determine the correction that should be applied to both
experiments.

\begin{figure}
\begin{center}
\vskip -0.2cm
\includegraphics[angle=0,width=3.7in]{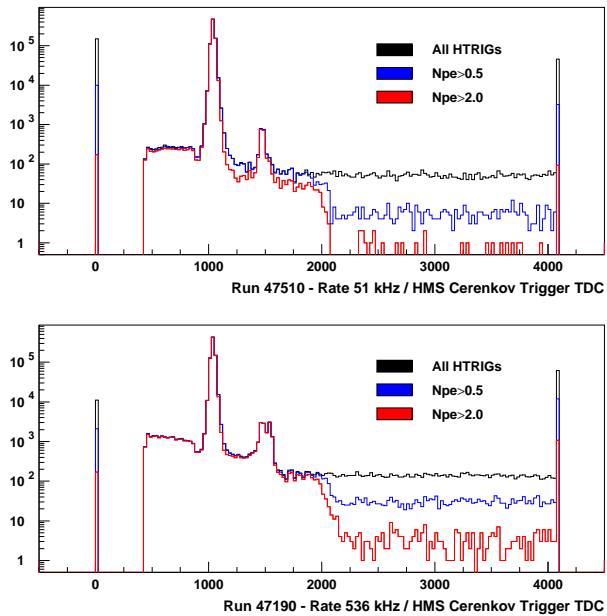}
\vskip -0.2cm
\caption{\label{fig:Cerenkov_TDC_2runs}
(Color online) HMS \u{C}erenkov Trigger multi-hit TDC histograms for two
  F$_{\pi}$-2 runs with \u{C}erenkov veto disabled.  The
  top panel is a low rate run, and the bottom panel is a high rate run.  HMS
  singles events, subject to a variety of indicated \u{C}erenkov cuts, are used
  in both spectra.  The TDC scale is 100~ps/chan.  Please see the text for
  further information.}
\end{center}
\end{figure}

In our first study, the timing spectra features of the \u{C}erenkov signal into
the HMS trigger were investigated for a variety of F$_{\pi}$-2 $\pi^-$ runs
with HMS singles rates between 7~kHz and $\sim$600~kHz.  The multi-hit TDC
is started by the HMS pretrigger signal and can be stopped multiple times by
the retimed (i.e. delayed and discriminated) \u{C}erenkov signal
(Fig. \ref{fig:Cerenkov_TDC_2runs}).  The main peak corresponds to signals
(primarily electrons) that result in the trigger, starting the TDC.  Events not
associated with the original trigger (other electrons or pions) appear as
additional events to the left and right of the main electron peak.  The second
peak to the right is due to a second electron arriving within the timing
window, but after the discriminator ``dead window'' of $\sim$40~ns (caused
by the length of the discriminator pulse).  The backgrounds to the left and
right of the two peaks are due to earlier and later electrons, while the tail
extending to channel 4096 is due to pedestal noise that crosses the
discriminator threshold.  The peak at channel 4096 is the accumulation of very
late TDC stops, while zeros correspond to electrons (or pions) that did not
give a stop.

As indicated by the differences between the low rate and high rate runs plotted
in Fig.~\ref{fig:Cerenkov_TDC_2runs}, the main peak to pedestal ratio degrades
with increasing rate, and the second peak to first peak ratio gets larger.  The
width of the portion of the TDC spectrum corresponding to electrons traversing
the detector current-to or after the original trigger particle indicated
that the effective \u{C}erenkov TDC gate width was 116.4$\pm$ 6.3 ns for the
F$_{\pi}$-2 $\pi^-$ runs, where the uncertainty is estimated from the slopes
and widths of the TDC spectra features.  We confirmed that the basic features
of the TDC spectra are the same for HMS singles and HMS+SOS coincidences.  We
also compared the TDC spectra for five pairs of $\pi^-$ runs, where for each
pair the beam and rate conditions were identical but in one run the HMS
\u{C}erenkov veto was disabled and in the other it was enabled.  The spectra
for runs with \u{C}erenkov trigger veto had a much greater proportion of events
where no TDC stop was recorded, due to the \u{C}erenkov signal being below the
discriminator threshold.  From the normalized differences of these pairs of
runs we estimated that the \u{C}erenkov trigger was about 90\% efficient at
vetoing electrons.

A comparison of $\pi^-$ runs with same rate but different trigger
condition can also be used to determine the effective threshold of the
\u{C}erenkov trigger veto.  The normalized difference of \u{C}erenkov
photoelectron (ADC) spectra was formed for each pair of runs, and the excess of
counts at a large number of photoelectrons when the veto was disabled indicated
an effective veto threshold of approximately 2.5 photoelectrons.  Because PMT
gain variations and pile-up effects will cause the actual veto threshold to
vary with rate, a slightly more restrictive software threshold on the number of
photoelectrons detected in the HMS \u{C}erenkov, $hcer_{\text{npe}}<2.0$, was
uniformly applied in the F$_{\pi}$-2 data analysis to cut out electrons.

In our second study, we made use of the same dedicated F$_{\pi}$-2
$\pi^-$ runs already used to determine the liquid deuterium cryotarget boiling
correction.  The \u{C}erenkov veto was disabled in all of these runs, and the
beam current was varied over a wide range for each $\pi^-$ kinematic setting
except for the high $\epsilon$ setting at \qsq=2.45~\gevsq, $E_e=5.25$~GeV,
$\theta_{\text{HMS}}=13.61^{\circ}$.  HMS fiducial and $hcer_{\text{npe}}<2.0$
cuts were applied to these HMS singles data, and the normalized $\pi^-$ yields
(with HMS tracking efficiency corrected by Eqn.~\ref{eqn:tracking_corr}) were
plotted versus HMS electron rate.  The normalized pion yield is expected to
drop with rate because of electrons passing through the \u{C}erenkov detector
within the trigger gate width after a pion has traversed the detector.  The
rate-dependences of the normalized pion yields at each kinematic setting were
consistent within their (large) uncertainties, and yielded an average gate
width of $\tau=139\pm 19$~ns.  Note that this study depends upon the tracking
efficiency and cryotarget boiling corrections used, while the first study based
on the \u{C}erenkov TDC spectra does not.  Finally, since the $\tau$ value from
the second study was determined with singles events, it needs to be adjusted to
yield the effective gate width for coincidence events.  This correction is
determined from the portion of the \u{C}erenkov TDC spectrum corresponding to
early electrons passing through the detector before the particle associated
with the trigger, yielding $99.2\pm 19$~ns.

The two F$_{\pi}$-2 \u{C}erenkov blocking studies (TDC gate width of $ 116.4\pm
6.3$ ns and corrected singles value of $99.2\pm 19$ ns) are consistent within
uncertainties.  It is difficult to tell which one is more definitive, so the
error weighted average $\tau_{\text{eff}}=114.7\pm 6.0$ ns, is used to compute
the \u{C}erenkov blocking correction
$\delta_{\text{CCblock}}=e^{-(\text{ELECTRONrate})\cdot\tau_\text{{eff}}}$ for the
F$_{\pi}$-2 $\pi^-$ analysis.  For the F$_{\pi}$-2 $\pi^-$ data, the HMS
electron rate varied from nearly zero to $\sim$600~kHz, resulting in a
\u{C}erenkov blocking correction of 0-6\%.  The $\pm 6.0$ ns uncertainty gives
an uncorrelated systematic uncertainty of 0.3\% at 500~kHz, while the 17 ns
difference in $\tau$ values from the two methods gives a scale uncertainty of
0.8\%.

$\pi^-$ data without \u{C}erenkov veto at different rates were unfortunately
not taken during the F$_{\pi}$-1 experiment, so the \u{C}erenkov blocking
correction cannot be directly determined for those data.  We therefore modify
the \u{C}erenkov blocking correction determined from F$_{\pi}$-2 data for use
in the F$_{\pi}$-1 analysis according to the following procedure.  A HMS
\u{C}erenkov photoelectron histogram for a carbon elastics run taken at the
very beginning of F$_{\pi}$-1, immediately before the first $\pi$ data run,
indicates that the effective veto threshold in the F$_{\pi}$-1 experiment is
slightly lower than that used in F$_{\pi}$-2.  Therefore, a slightly more
restrictive software threshold of $hcer_{\text{npe}}<1.5$ was applied in the
analysis of the F$_{\pi}$-1 data.  The figure also indicates that the
\u{C}erenkov veto would be about 80\% efficient for this run.

\begin{figure}
\begin{center}
\vskip -0.2cm
\includegraphics[angle=0,width=3.65in]{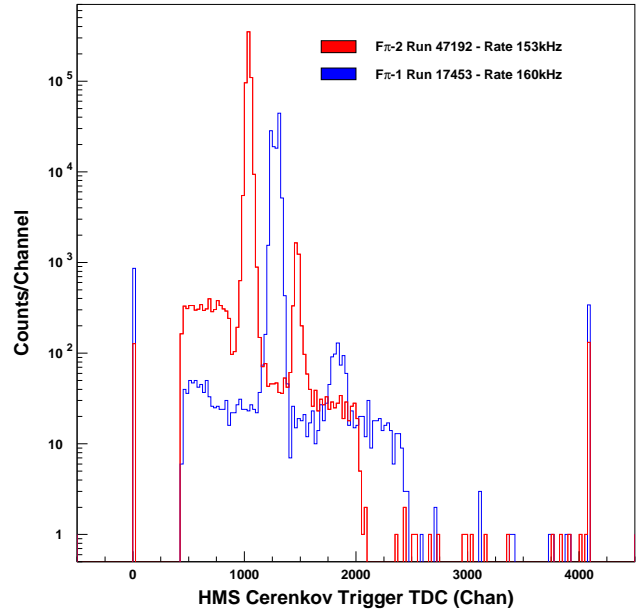}
\vskip -0.2cm
\caption{\label{fig:Cerenkov TDC}
(Color online) HMS \u{C}erenkov Trigger TDC histogram for the one F$_{\pi}$-1
  $\pi^-$ run with \u{C}erenkov veto disabled (blue), compared to a
  F$_{\pi}$-2 run with same trigger at similar rate (red).  HMS singles events,
  subject to a $hcer_{\text{npe}}>2.0$ \u{C}erenkov cut, are used in both
  spectra.  The TDC scale is 100~ps/chan.}
\end{center}
\end{figure}

We therefore reanalyzed the F$_{\pi}$-2 dedicated $\pi^-$ runs without
\u{C}erenkov veto, except that a $hcer_{\text{npe}}<1.5$ \u{C}erenkov particle
identification appropriate to the F$_{\pi}$-1 analysis was applied.  The
dependence of normalized pion singles yields on rate yielded a value of
$\tau=162\pm 19$ ns, which was then adjusted to give an effective gate width
for coincidence events of $116\pm 20$ ns.  Finally, we used the TDC timing
information from the only F$_{\pi}$-1 ``open trigger'' run taken just before
the main data taking to estimate the scaling with respect to the F$_{\pi}$-2
timing information.  As shown in Fig. \ref{fig:Cerenkov TDC}, the TDC timing
window used during F$_{\pi}$-1 is wider than in F$_{\pi}$-2.
Comparing the equivalent features of the two spectra gives a scale factor of
$1.19\pm 0.084$.  Application of this scale factor to the $\tau$ value
determined from the F$_{\pi}$-2 data yields $\tau=(115.7\pm 20)\times (1.19\pm
0.084)=137.7\pm 26$ ns.

The two values compare well (TDC gate width of $ 138.4\pm 6.3$ ns and corrected
singles value of $137.7\pm 26$ ns) and thus the error-weighted average
$\tau_{\text{eff}}=138.4\pm 6.1$ ns of the two was taken as the effective
$\tau$ value to compute the \u{C}erenkov blocking corrections for the
F$_{\pi}$-1 data normalization.  For the F$_{\pi}$-1 $\pi^-$ data, the HMS
electron rate varied from nearly zero to $\sim$1.2~MHz, resulting in a
\u{C}erenkov blocking correction of 0-15\%.  The $\pm 6.1$ ns uncertainty gives
an uncorrelated systematic uncertainty of 0.7\% at 1.2 MHz, and scaling the
0.8\% F$_{\pi}$-2 scale uncertainty to 1.2 MHz gives a scale uncertainty of
1.0\%.

\subsection{Other Particle Identification Corrections
\label{sec:betacorr}}

\begin{figure}
\vskip -0.3cm 
\begin{center}
\includegraphics[angle=0,width=3.8in]{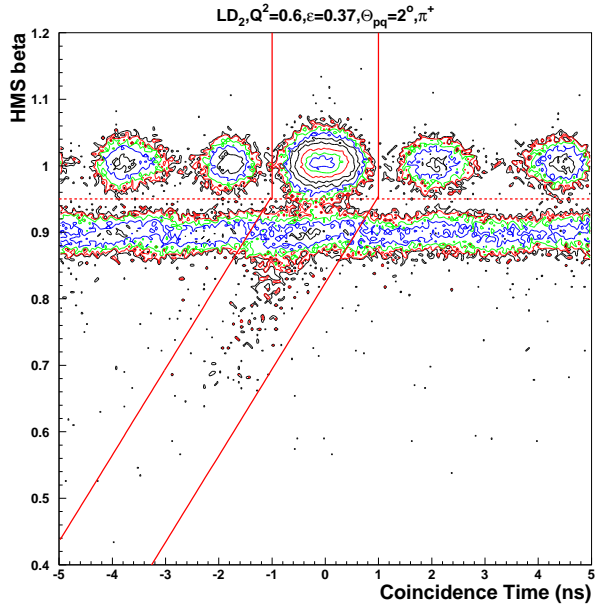}
\vskip -0.3cm
\caption{\label{fig:beta_HMS_for_Fpi-1_data_set} 
(Color online) HMS+SOS coincindence time versus $\beta_{\text{HMS}}$ for a
representative F$_{\pi}$-1 $\pi^+$ run.  The dashed line indicates the $\beta
>0.95$ cut used to separate pions from protons.  The solid lines indicate the
region (for $\pi^-$ runs, without proton contamination) used to compute the
$\beta$ cut correction.  See the text for more details.}
\end{center}
\end{figure}

Fig. \ref{fig:beta_HMS_for_Fpi-1_data_set} shows the HMS particle speed,
$\beta=v/c$, which is calculated from the time-of-flight difference between two
scintillator planes in the HMS detector stack.  The upper band events are
$\pi^+$ in the HMS, with the 2~ns beam structure of the incident electron beam
clearly visible.  The lower band events are protons.  In both F$_{\pi}$-1 and
F$_{\pi}$-2, a cut $\beta>0.95$ was used to eliminate the protons.
Additionally in the F$_{\pi}$-2 experiment, an aerogel \u{C}erenkov detector
was used for separating protons and $\pi^+$ for HMS central momenta above 3
GeV/c.

Figure \ref{fig:beta_HMS_for_Fpi-1_data_set} also displays a ``tail'' at low
${\beta_{\text{HMS}}}$ due to pions undergoing nuclear interactions in the
scintillators, \u{C}erenkov detector material, and in the case of F$_{\pi}$-2
experiment, the aerogel \u{C}erenkov detector material.  A correction for pion
events at lower $\beta$ eliminated by the $\beta>0.95$ cut was applied.  In
F$_{\pi}$-1 this correction was extracted from the $\pi^-$ data and was applied
to both the $\pi^-$ and $\pi^+$ data sets.  The correction was 4.89\%, with an
uncertainty of 0.41\% determined from the standard deviation of the correction
determined from the different $\pi^-$ kinematic settings.  For the F$_{\pi}$-2
data, the same procedure was used, except that the aerogel \u{C}erenkov
detector permitted the separation of protons from pions, leading to a cleaner
pion sample.  For each $\pi^+$ and $\pi^-$ kinematic setting, ``beta cut
corrections'' were extracted in the same fashion, yielding average $\beta$ cut
corrections of $2.42\%\pm 0.12\%$ and $2.51\%\pm 0.18\%$ for $\pi^+$ and
$\pi^-$, respectively.

A correction for the number of pions lost due to pion nuclear interactions and
true absorption in the HMS exit window and detector stack of 1-2\% was also
applied.  For details on how this correction was determined, see
Ref.~\cite{Blok08}.

A comprehensive summary of the various corrections applied to the data is given
in Table \ref{tab:eff_summary}.

\begin{table*}
\begin{center}
\begin{tabular}{|l|c|c|}\hline
\multicolumn{3}{|c|}{Summary of F$_{\pi}$-1 Correction Factors}\\
\hline\hline
HMS tracking efficiency correction    & $1-(0.0676\pm 0.002)/$S1Xrate(MHz) &
Sec.~\ref{sec:trackeff}\\
LD$_2$ Cryotarget Boiling             & $1-(0.0472\pm 0.003)/100~\mu$A &
Sec.~\ref{sec:ld2_boiling}\\
Beam Current Offset                   & $0.2\mu$A & Sec.~\ref{sec:ld2_boiling}\\
HMS \u{C}erenkov blocking             & $e^{{\rm -(\text{ELECTRONrate})}\cdot(138.4\pm
  6.1~{\rm ns})}$ & Sec.~\ref{sec:cerblock}\\
$\beta_{cut}$ correction ($\pi^{\pm}$) & $4.89\% \pm 0.41\%$ &
Sec.~\ref{sec:betacorr}\\
Pion Absorption & $1\%\pm 1\%$ & Sec.~\ref{sec:betacorr}, Ref.~\cite{Blok08}\\
\hline
SOS \u{C}erenkov efficiency           & $99.92\%\pm 0.02\%$ & Ref.~\cite{Tanjathesis}\\
SOS Calorimeter efficiency            & $99.5\%\pm 0.1\%$ & Ref.~\cite{Tanjathesis}\\
HMS \u{C}erenkov efficiency           & $99.6\%\pm 0.05\%$ & Ref.~\cite{Tanjathesis}\\
Coincidence Time Blocking             & $e^{{\rm -Total Pretrig rate}\cdot(140
  ~{\rm ns})}$ & Ref.~\cite{Volmerthesis}\\ 
HMS electronic live time              & $1 - 5/6(N_{h60}-N_{h120})/N_{\text{hELREAL}}$ &
Ref.~\cite{Volmerthesis}\\
SOS electronic live time              & $1 - 5/6(N_{s60}-N_{s120})/N_{\text{sELREAL}}$ &
Ref.~\cite{Volmerthesis}\\

\hline\hline

\multicolumn{3}{|c|}{Summary of F$_{\pi}$-2 Correction Factors}\\
\hline\hline
HMS tracking efficiency correction    & $1-(0.0676\pm 0.002)/$S1Xrate(MHz) &
Sec.~\ref{sec:trackeff}\\
LD$_2$ Cryotarget Boiling             & No correction.  $\pm 0.3\%/100~\mu A$ &
Sec.~\ref{sec:ld2_boiling}\\
HMS \u{C}erenkov blocking             & $e^{{\rm -\text{ELECTRONrate}}\cdot(114.7\pm
  6.0~{\rm ns})}$ & Sec.~\ref{sec:cerblock}\\
$\beta_{cut}$ correction ($\pi^-$)     & $2.51\% \pm 0.18\%$ & Sec.~\ref{sec:betacorr}\\
$\beta_{cut}$ correction ($\pi^+$)     & $2.42\% \pm 0.12\%$ & Sec.~\ref{sec:betacorr}\\
Pion Absorption                       & $2\%\pm 1\%$ & Sec.~\ref{sec:betacorr}, Ref.~\cite{Blok08}\\
\hline
SOS \u{C}erenkov efficiency           & $99.92\%\pm 0.02\%$ & Ref.~\cite{Tanjathesis}\\
SOS Calorimeter efficiency            & $99.5\%\pm 0.1\%$ & Ref.~\cite{Tanjathesis}\\
HMS \u{C}erenkov efficiency           & $99.6\%\pm 0.05\%$ & Ref.~\cite{Tanjathesis}\\
HMS Aerogel efficiency                & $99.5\%\pm 0.02\%$ & Ref.~\cite{Tanjathesis}\\
Coincidence Time Blocking             & $e^{{\rm -SOS Pretrig rate}\cdot(92
  ~{\rm ns})}$ & Ref.~\cite{Tanjathesis}\\
HMS electronic live time              & $1 - 6/5(N_{h100}-N_{h150})/N_{h100}$ &
Ref.~\cite{Tanjathesis}\\
SOS electronic live time              & $1 - 6/5(N_{s100}-N_{s150})/N_{s100}$ &
Ref.~\cite{Tanjathesis}\\

\hline
\end{tabular}
\end{center}
\vskip -.5cm
\caption{Summary of corrections applied to the deuterium data.  In addition, 
  HMS and SOS tracking efficiencies and computer live times are applied on a
  run-by-run basis.  The electronic livetimes are measured by counting pretrigger
  signals with different gate widths $N_X$.
\label{tab:eff_summary}}
\end{table*}

\subsection{Backgrounds}

The coincidence timing structure between unrelated electrons and protons or
pions from any two beam bursts is peaked every 2~ns, due to the accelerator
timing structure.  Real and random $e$-$\pi$ coincidences were selected with a
coincidence time cut of $\pm 1$ ns.  The random coincidence background (2-10\%
during F$_{\pi}$-1, depending on the kinematic setting, 1-2\% during
F${_\pi}$-2) were subtracted on a bin by bin basis.

The contribution of background events from the aluminum cell walls was
estimated using dedicated runs with two ``dummy'' aluminium targets placed at
the appropriate locations. These data were analyzed in the same way as the
cryotarget data and the yields  (2-4\% of the total yield) were subtracted from
the cryotarget yields, taking into account the different thicknesses (about a
factor of seven) of the target-cell walls and dummy target.  The contribution
of the subtraction to the total uncertainty is negligible.

\section{Cross Section Determination and Systematic Uncertainties
\label{sec:analysis}}

\subsection{Method}

Following our earlier procedure \cite{Blok08}, we write the unpolarized pion
electroproduction cross section as the product of a virtual photon flux factor
and a virtual photon cross section,
\begin{equation}
   \frac{d^5 \sigma}{d \Omega_e dE_e^\prime d \Omega_{\pi}} = J\left(t,\phi
   \rightarrow \Omega_{\pi}\right) \Gamma_v \frac{d^2 \sigma}{dt d \phi},
\end{equation}
where $J\left(t,\phi \rightarrow \Omega_{\pi}\right)$ is the Jacobian of the
transformation from $dtd\phi$ to $d\Omega_{\pi}$, $\phi$ is the azimuthal
angle
between the scattering and the reaction plane, and $\Gamma_v$=$\frac{\alpha}{2
\pi^2} \frac{E^\prime_e}{E_e} \frac{1}{Q^2} \frac{1}{1-\epsilon}
\frac{W^2-M^2}{2 M}$ is the virtual photon flux factor. 

The (reduced)
cross section can be expressed in terms of contributions from transversely and
longitudinally polarized photons,
\begin{eqnarray}
\label{eqn:unsep}
  2\pi \frac{d^2 \sigma}{dt d\phi} & = & \frac{d \sigma_T}{dt} + \epsilon
  \frac{d \sigma_L}{dt} + \sqrt{2 \epsilon (1 + \epsilon)} \frac{d
  \sigma_{LT}}{dt} \cos \phi \\ 
  \nonumber & + & \epsilon \frac{d \sigma_{TT}}{dt} \cos 2 \phi.
\end{eqnarray}
Here,
$\epsilon=\left(1+2\frac{|\vec{q}|^2}{Q^2}\tan^2\frac{\theta}{2}\right)^{-1}$
is the virtual photon polarization, where $\vec{q}$ is the
three-momentum transferred to the quasi-free nucleon, $\theta$ is the
electron scattering angle, and $\phi$ has already been defined.

In order to separate the different structure functions, one has to determine
the cross section both at high and at low $\epsilon$ as a function of 
$\phi$ for fixed values of $W$, \qsq\ and $t$. Since the $t$-dependence is
important, this should be done for various values of $t$ at every central \qsq\
setting.  Therefore, the data are binned in $t$ and $\phi$, thus integrating
over $W$ and \qsq\ within the experimental acceptance, and also over
$\theta_\pi$ (the latter is of relevance, since the interference structure
functions include a dependence on $\sin \theta_\pi$).  However, the average
values of $W$, \qsq, and $\theta_\pi$ generally are not the same for different
$\phi$ and for low and high $\epsilon$. Moreover, the average values of $W$,
\qsq, $t$, and $\theta_\pi$, only three of which are independent, may be
inconsistent.

Both problems can be avoided by comparing the measured yields to
the results of a Monte-Carlo simulation for the actual experimental
setup, in which a realistic model of the cross section is implemented.
At the same time, effects of finite experimental resolution, pion decay,
radiative effects, etc., can be taken into account.
When the model describes the dependence of the four structure functions on $W$,
\qsq, $t$, $\theta_\pi$ sufficiently well, i.e. when the ratio of 
experimental to simulated yields is close to unity within the statistical
uncertainty, the cross section for any value of 
$\overline{W}$, $\overline{Q^2}$ within the acceptance can be determined as 
\begin{equation}
\label{eqn:ratio_to_sigma}
\left( \frac{d^2 \sigma}{dt d\phi}(t,\phi) \right)^{\mathrm{exp}}_{\overline{W},\overline{Q^2}}
=\frac{Y_{\mathrm{exp}}}{Y_{\mathrm{sim}}}
\left( \frac{d^2 \sigma}{dt d\phi}(t,\phi) \right)^{\mathrm{model}}_{\overline{W},\overline{Q^2}},
\end{equation}
where $Y$ is the yield over $W$ and \qsq, with common values of
$\overline{W}$, $\overline{Q^2}$ (if needed different for different values of $t$)
for all values of $\phi$, and for the high and low $\epsilon$
data, so as to enable a separation of the structure functions.  In practice the
data at both high and low $\epsilon$ were binned in 4-6 $t$-bins and 16
$\phi$-bins and the cross section was evaluated at the center of each bin. The
overlined values in the expression above were taken as the acceptance weighted
average values for all $\phi$-bins (at both high and low $\epsilon$) together,
which results in them being slightly different for the $t$-bins.

\subsection{Description of the Simulation Model and Kinematic Variables
\label{sec:model}}

The Hall C Monte Carlo package, SIMC, is used as the simulation package for this
experiment.  A detailed description of the program is given in
Refs. \cite{Blok08,Volmerthesis, Gaskellthesis}.
For each event, the program generates the coordinates of the
interaction vertex ($x,y,z$) and the three-momenta of the scattered electron
and the produced pion for the $^2$H$(e,e'\pi^{\pm})NN_s$ reaction.
In the SIMC event generator, the following off-shell prescription was taken to
determine the kinematics.  The ``spectator'' nucleon was taken to be
``on-shell'' in the initial state, while the struck nucleon was taken to be
``off-shell'' with the requirement that the total momentum of the nucleus is
zero, and the total energy is the mass of a deuteron, $M_{D}$. The nucleon on
which the pion is produced thus has a certain momentum (Fermi motion), taken
from a deuteron wave function calculated with the Bonn $NN$-potential
\cite{Koltenukthesis}.  The outgoing particles are followed on their way
through the target, spectrometer and detector stack, taking into account energy
loss and multiple scattering. Possible radiation of the incoming and outgoing
electron and the pion is included \cite{Blok08,ent00}.  This leads to
`experimental' values for the three-momenta of the scattered electron and the
produced pion.  Together with the value for the incoming electron, these are
used to calculate kinematic quantities such as \qsq, $W$, $t$, $\theta_\pi$,
and $\phi_\pi$, just as for the experimental data.

\begin{figure}
\begin{center}
\includegraphics[width=3.25in]{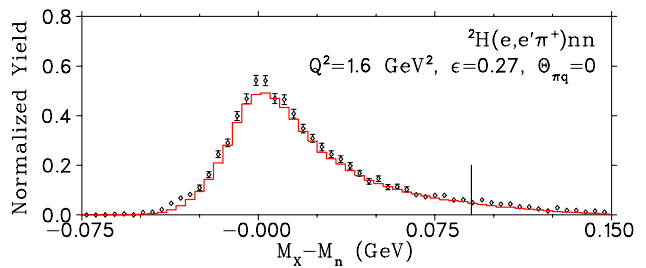}
\end{center}
\caption{(Color online) 
  Missing mass of the undetected nucleon calculated as
  quasi-free pion electroproduction for a representative $\pi^+$ setting.  
  The diamonds are experimental data, and the red line is the quasi-free Monte
  Carlo simulation.  The vertical line indicates the $M_X$ cut upper limit.}
\label{fig:MMplot}
\end{figure}

Because experimentally the momentum of the struck nucleon is not observable,
the kinematic quantities $t$, missing mass $M_X$, and $\theta_\pi$ were
reconstructed (both for the experimental data and for the SIMC data) assuming
quasi-free pion electroproduction, $\gamma^* N \rightarrow \pi^{\pm} N'$, where
the virtual photon interacts with a nucleon at rest.  The Mandelstam variable
$t$ is calculated as $t=(p_{\text{target}}-p_{\text{recoil}})^2$.  (In the
limit of perfect resolution and no radiative effects, for $^1$H this formula
gives the same result as $(p_{\gamma}-p_{\pi})^2$, but for $^2$H it does not,
because of binding effects.)  The missing mass $M_X$ was calculated according
to:
\begin{eqnarray}
  \vec{p}_{\text{missing}}&=&\vec{q}-\vec{p}_{\pi},\nonumber\\ 
  E_{\text{missing}}&=&\nu+m_N-E_{\pi},\\
  M_X^2&=&E_{\text{missing}}^2-p_{\text{missing}}^2\nonumber
\end{eqnarray}
where $m_N$ equals the free proton mass for $\pi^+$ production and the free
neutron mass for $\pi^-$ production.  See Fig. \ref{fig:MMplot} for a
representative example.
Finally, the center of mass (CM) frame azimuthal angle $\phi_{\text{CM}}$ in
Eqn.~\ref{eqn:unsep} equals the experimentally reconstructed $\phi_{\pi q}$ and
$\theta_{\text{CM}}$ is calculated by boosting to the photon plus nucleon at
rest system.


Event weighting in the simulation used a model cross section that
depends on the values of \qsq, $W$, $t$, $\theta_\pi$, and $\phi_\pi$,
calculated in the same way as for the (experimental and simulated) data, but
using the vertex $k_{e'}$ and $k_{\pi}$.  
An iterative fitting procedure, discussed in Sec.~\ref{sec:modelxsec}, was 
used to determine this model cross section.

It should be stressed that because of the quasi-free assumption with an initial
nucleon at rest, the extracted cross sections and structure functions are
effective ones, which cannot be directly compared to those from $^1$H.  It was
considered better that the influence of off-shell effects (and possible other
mechanisms in $^2$H) are studied separately, using cross sections that were
determined in a well defined way, than that off-shell effects are incorporated
already in some way in the extracted cross sections.  (Although the differences
in practice may not be large.)

In extracting the deuterium cross sections, it is desirable to keep as much of
the missing-mass tail as possible (up to the two-pion threshold of 1.1 GeV), to
maximize the acceptance of the ``quasifree'' distribution, and to minimize the
systematic uncertainty associated with the missing mass cut. 

The thick collimators of the HMS and SOS are very effective at stopping
electrons, but a non-negligible fraction of the pions undergo multiple
scattering and ionization energy loss and consequentially end up contributing
to the experimental yield~\cite{Gaskellthesis}.  These pion (hadron)
punch-through events have been observed in earlier experiments, and corrections
are needed for a precise yield extraction.  Since the pions in F$_{\pi}$-1 and
F$_{\pi}$-2 are detected in the HMS, the implementation of the simulated
collimator punch-through events was done for only this arm.  The HMS event
simulation therefore takes into account the probability that a pion interacts
hadronically with the collimator (allowing the pion to undergo multiple
scattering and ionization energy loss).  After implementing the pion
punch-through events in SIMC, the $M_X$ cut upper limit was determined by the
value where the missing mass peak is no longer well reproduced by a quasi-free
Monte Carlo simulation including all known detector effects, indicating the
presence of additional backgrounds, such as two pion production.  The missing
mass cut was taken to be 0.875$\leq M_X\leq$1.03 GeV.  It is wider than the one
used in the analysis of the hydrogen data because of Fermi motion in the
deuteron.  Compared to hydrogen, the backgrounds from target windows and random
coincidences are generally larger due to the wider $M_X$ cut.

\subsection{Determination of Separated Structure Functions}
\label{sec:modelxsec}

\begin{figure}
\begin{center}
\includegraphics[angle=0,width=3.0in]{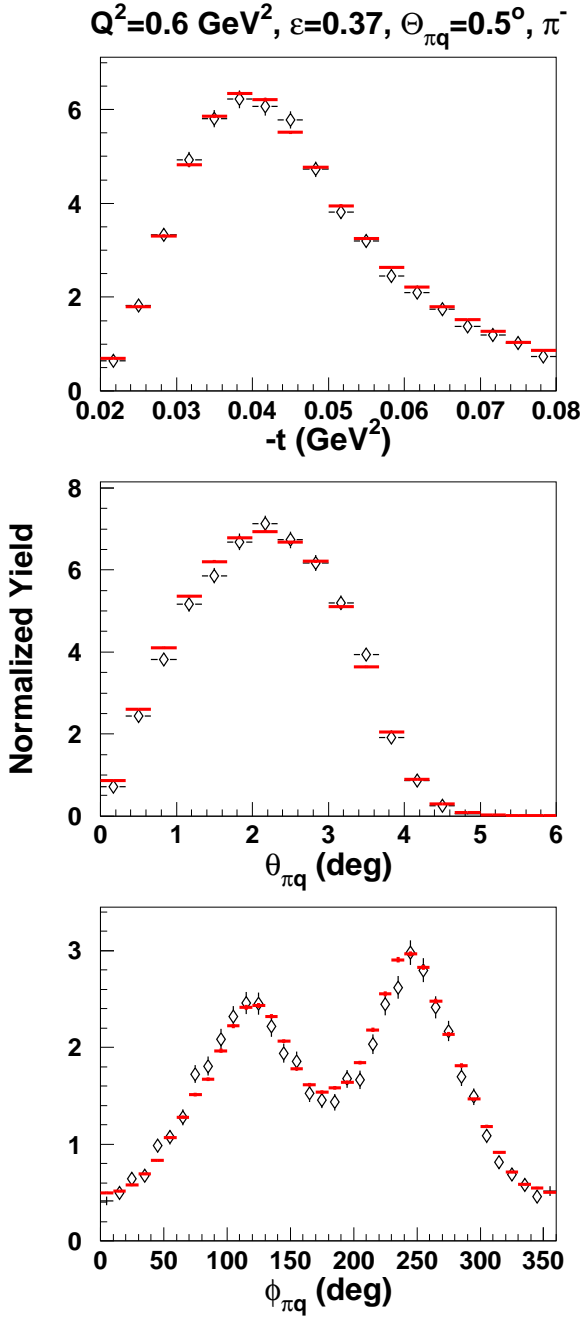}
\caption{\label{fig:ExpKin}
(Color online) Normalized experimental $\pi^-$ yield (black
    diamonds) in comparison to the quasi-free Monte Carlo simulation (red
    lines) for one HMS+SOS setting at \qsq=0.60 \gevsq, low $\epsilon$.}
\end{center}
\end{figure}

\begin{figure}
\begin{center}
\includegraphics[angle=0,width=3.0in]{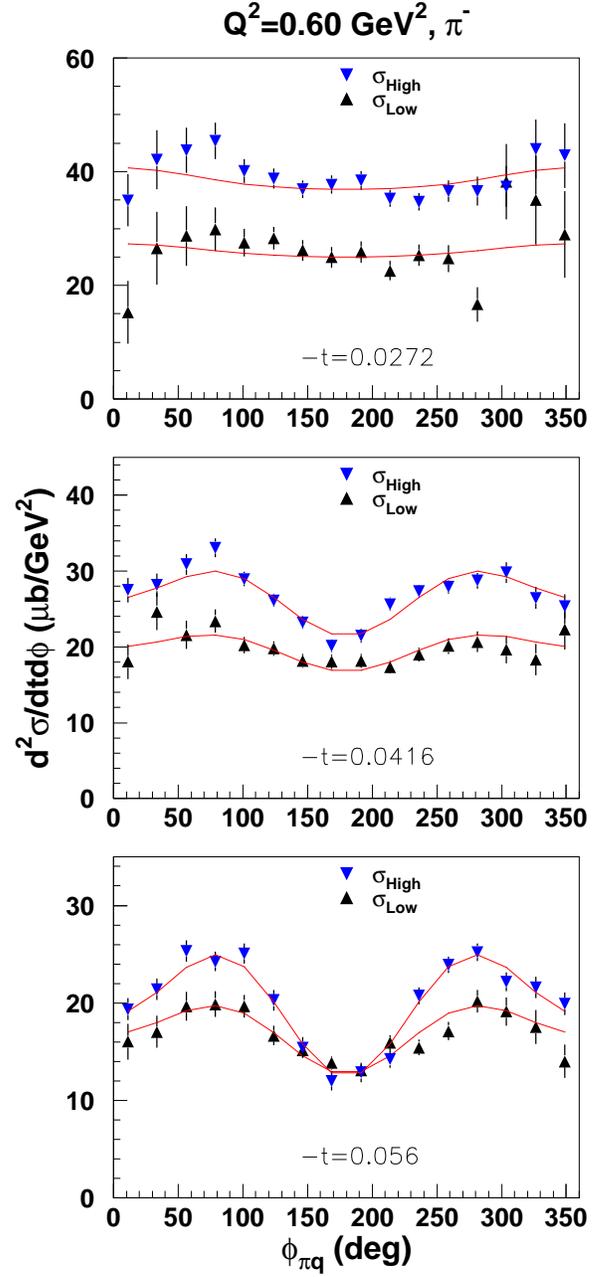}
\caption{\label{fig:Phifit}
(Color online) Unseparated experimental $\pi^-$ cross sections as a function of
azimuthal angle $\phi$ at \qsq=0.60 \gevsq, low $\epsilon$ (black triangles)
and high $\epsilon$ (blue inverted triangles).  The
curves shown represent the fit of the measured values of the cross section to
Eqn.~\ref{eqn:unsep}.  In this fit, all four parameters $\sigma_{L, T, LT, TT}$
are extracted simultaneously, separate for each $-t$ bin.  }
\end{center}
\end{figure}

The SIMC model cross section and the final separated structure functions were
determined in the same (iterative) procedure. The model cross section was taken
as the product of a global function describing the $W$-dependence times (a sum
of) \qsq\ and $t$ dependent functions for the different structure functions.
For the $LT$ and $TT$ parts, their leading order dependence on
$\sin$($\theta_{CM}$) was taken into account \cite{raskin}. The $W$-dependence
was taken as $(W^2-M_N^2)^{-2}$, where $M_N$ is the struck nucleon mass, based
on analyses of experimental data from Refs.~\cite{Brauel1,beb78}. For the parts
depending on \qsq\ and $t$, phenomenological forms were used and the parameters
were fitted. For all $t$-bins at every (central) \qsq\ setting,
$\phi$-dependent cross sections were determined both at high and low $\epsilon$
for chosen values of $\overline{W}$, $\overline{Q}^2$ (and corresponding values
of $\theta_{\pi}$ and $\epsilon$) according to Eqn.~\ref{eqn:ratio_to_sigma}.
The iteration procedure was repeated until satisfactory agreement between the
experimental and simulated distributions was obtained, the values of
$\sigma_{L,T,LT,TT}$ (and the associated fit parameters) were stable in
subsequent iterations, and the parameters fitted at the individual
\qsq-settings did not change much with \qsq.  A representative example of 
some relevant variables and of the fit of the
experimental cross section as a function of $\phi_\pi$ is shown in
Figs.~\ref{fig:ExpKin},~\ref{fig:Phifit}.  The cosine structure from the
interference terms is clearly visible in Fig.~\ref{fig:Phifit}.

This procedure was carried out independently for $\pi^+$ and $\pi^-$ at each
\qsq, in order to have optimal descriptions in the different kinematic
ranges covered.  The parameterizations used in the F$_{\pi}$-1 $\pi^+$ analysis
are:
\begin{eqnarray}
\frac{d\sigma_{L}}{dt} &=& g(W) \bigl(p_{1}+p_{2} \ln(Q^2)\bigr)
e^{(p_{3}+p_{4}\ln(Q^2))(-t)},\nonumber\\
\frac{d\sigma_{T}}{dt} &=& g(W) 
\biggl(\frac{|t|-|t_{\text{ave}}|}{|t_{\text{ave}}|}\biggr)\nonumber\\
&\times&\biggl( p_{5}+p_{6}\ln(Q^2)
+\bigl(p_{7}+p_{8}\ln(Q^2)\bigr)\biggr),\nonumber\\
\frac{d\sigma_{LT}}{dt} &=& g(W) p_{9} e^{p_{10}(-t)} \sin\theta_{\text{CM}},\\
\frac{d\sigma_{TT}}{dt} &=& g(W) f(t) 
\frac{p_{11}}{Q^2 }e^{-Q^2} \sin^2\theta_{\text{CM}},\nonumber
\label{eqn:mc_model_pl}
\end{eqnarray}
where $g(W)=1/(W^2-m_p^2)^2$ is the assumed $W$-dependence discussed earlier,
$f(t)=-t/(-t-m_{\pi}^2)^2$ is the pion pole factor, $|t_{\text{ave}}|$ is the
average $-t$ value for a given kinematic setting, given by
$|t_{\text{ave}}|=(0.105+0.04 \ln(Q^2))Q^2$, and $p_{i=1,...,12}$ are the fit
parameters.

For the F$_{\pi}$-1 $\pi^-$ analysis, a slightly different parameterization
(because \sigt\ and \sigtt\ showed a stronger 
\qsq-dependence)  yielded a better fit:
\begin{eqnarray}
\frac{d\sigma_{L}}{dt} &=& g(W) \bigl(p_{1}+p_{2} \ln(Q^2)\bigr)
e^{(p_{3}+p_{4}\ln(Q^2))(-t)},\nonumber \\
\frac{d\sigma_{T}}{dt} &=& g(W) \biggl( p_{5}+\frac{p_{6}}{Q^4+0.1}\nonumber\\
&+&\bigl(p_{7}+p_{8}\ln(Q^2)\bigr)\biggl(
\frac{|t|-|t_{\text{ave}}|}{|t_{\text{ave}}|}\biggr)\biggr),\nonumber \\
\frac{d\sigma_{LT}}{dt} &=& g(W) p_{9} e^{p_{10}(-t)} \sin\theta_{\text{CM}},\\
\frac{d\sigma_{TT}}{dt} &=& g(W) f(t) \biggl(
\frac{p_{11}}{Q^2 }+\frac{p_{12}}{Q^4+0.2}\biggr) \sin^2\theta_{\text{CM}}. 
\nonumber
\label{eqn:mc_model_mn}
\end{eqnarray}

In the F$_{\pi}$-2 analyses, a common parameterization (similar to those in
F$_{\pi}$-1) was used for both $\pi^+$ and $\pi^-$:
\begin{eqnarray}
\frac{d\sigma_{L}}{dt} &=& g(W) \bigl(p_{1}+p_{2} \ln(Q^2)\bigr)
e^{(p_{3}+p_{4}\ln(Q^2))(-t-0.2)},\nonumber \\
\frac{d\sigma_{T}}{dt} &=& g(W) \biggl( p_{5}+p_{6}\ln(Q^2)\nonumber\\
&+&\bigl(p_{7}+p_{8}\ln(Q^2)\bigr)\biggl(
\frac{|t|-|t_{\text{ave}}|}{|t_{\text{ave}}|}\biggr)\biggr),\nonumber \\
\frac{d\sigma_{LT}}{dt} &=& g(W) \biggl( p_{9} e^{p_{10}(-t)} + \frac{p_{11}}{(-t)}
\biggr) \sin\theta_{\text{CM}},\\
\frac{d\sigma_{TT}}{dt} &=& g(W) f(t) 
\frac{p_{12}}{Q^2 }e^{-Q^2} \sin^2\theta_{\text{CM}}, \nonumber
\label{eqn:mc_model_fpi2}
\end{eqnarray}
where $|t_{\text{ave}}|=\bigl(0.0735+0.028 \ln(Q^2)\bigr)Q^2$ and $p_4=0$.

\begin{table}
\begin{center}
\begin{tabular}{|l|c|c|c|}\hline 
Correction              &  Uncorrelated & $\epsilon$ uncorr. &  Correlated  \\
                        &  (pt-to-pt)   & {\it t} corr.      &  (scale)     \\
                        & [\%]          & [\%]               &  [\%]    \\ 
\hline \hline
 d$\theta_{e}$              &    0.1     &     0.7-1.1        &          \\
 d$E_{\text{beam}}$          &    0.1     &     0.2-0.3        &          \\
 d$P_{e}$                   &    0.1     &     0.1-0.3        &          \\ 
 d$\theta_{\pi}$            &    0.1     &     0.2-0.3        &          \\
 Radiative corr            &            &     0.4            &  2.0     \\
 HMS  $\beta$ corr         &    0.4     &                    &          \\
 Particle ID               &            &     0.2            &          \\ 
 Pion absorption           &            &                    &  1.0     \\
 Pion decay                &    0.03    &                    &  1.0     \\
 HMS Tracking ($\pi^+$)    &            &     0.4            &  1.0     \\
 HMS Tracking ($\pi^-$)    &            &     1.3            &  1.0     \\
 SOS Tracking              &            &     0.2            &  0.5     \\
 Charge                    &            &     0.3            &  0.5     \\
 Target Thickness          &            &     0.3            &  1.0     \\
 CPU dead time             &            &     0.2            &          \\
 HMS Trigger               &            &     0.1            &          \\
 SOS Trigger               &            &     0.1            &          \\
 Electronic DT             &            &     0.1            &          \\
 HMS Cer. block. ($\pi^-$) &    0.7     &                    &  1.0     \\
 Acceptance                &    1.0     &     0.6            &  1.0     \\
\hline\hline
 Total ($\pi^+$)           &    1.1     &     1.3-1.6        &  3.1  \\
 Total ($\pi^-$)           &    1.3     &     1.8-2.0        &  3.2  \\ 
\hline
\end{tabular}
\end{center}
\vskip -.5cm
\caption{\label{table:Fpi1_syst_unc_pl}
Systematic uncertainties for F$_{\pi}$-1.  Those items not discussed explicitly
in preceeding sections are assumed to be the same as for the published $^1$H
analysis.  These are the uncertainties in: kinematic offsets, radiative
corrections, pion decay, SOS tracking, trigger efficiency, CPU and electronic
dead time, and acceptance.  The systematic uncertainties in each column are
added quadratically to obtain the total systematic uncertainty.}
\end{table}

\begin{table}
\begin{center}
\begin{tabular}{|l|c|c|c|}\hline 
Correction              &  Uncorrelated & $\epsilon$ uncorr. &  Correlated  \\
                        &  (pt-to-pt)   & {\it t} corr.      &  (scale)     \\
                        & [\%]          & [\%]               &  [\%]    \\ 
\hline \hline
 d$\theta_{e}$             &    0.1     &     0.7-1.1        &          \\
 d$E_{\text{beam}}$         &    0.1     &     0.2-0.3        &          \\
 d$P_{e}$                  &    0.1     &     0.1-0.3        &          \\ 
 d$\theta_{\pi}$           &    0.1     &     0.2-0.3        &          \\
 Radiative corr            &            &     0.4            &  2.0     \\
 HMS $\beta$ corr ($\pi^+$)&    0.12    &                    &          \\
 HMS $\beta$ corr ($\pi^-$)&    0.18    &                    &          \\
 Particle ID               &            &     0.2            &          \\ 
 Pion absorption           &            &                    &  1.0     \\
 Pion decay                &    0.03    &                    &  1.0     \\
 HMS Tracking ($\pi^+$)    &            &     0.3            &  0.5     \\
 HMS Tracking ($\pi^-$)    &            &     0.45           &  0.75    \\
 SOS Tracking              &            &     0.2            &  0.5     \\
 Charge                    &            &     0.3            &  0.5     \\
 Target Thickness          &            &     0.2            &  0.8     \\
 CPU dead time             &            &     0.2            &          \\
 HMS Trigger               &            &     0.1            &          \\
 SOS Trigger               &            &     0.1            &          \\
 Electronic DT             &            &     0.1            &          \\
 HMS Cer. block. ($\pi^-$) &    0.3     &                    &  0.8     \\
 Acceptance                &    0.6     &     0.6            &  1.0     \\
\hline\hline
 Total ($\pi^+$)           &  0.6       &     1.2-1.5        &  2.9     \\
 Total ($\pi^-$)           &  0.7       &     1.3-1.6        &  3.1     \\ 
\hline
\end{tabular}
\end{center}
\vskip -.5cm
\caption{\label{table:Fpi2_syst_unc_pl}
Systematic uncertainties for F$_{\pi}$-2, similar to Table
\ref{table:Fpi1_syst_unc_pl}.  Those items not discussed explicitly in
preceeding sections are assumed to be the same as for the published $^1$H
analysis.}
\end{table}

\subsection{Systematic Uncertainties due to Missing
  Mass Cut and SIMC Model Dependence}

Since the extracted separated cross sections depend in principle on the cross
section model, there is a ``model systematic uncertainty.''  This uncertainty
was studied by extracting \sigl\ and \sigt\ with different cross section
models.  There is a second, related uncertainty due to the modeling of the
missing mass distribution.  The combined systematic uncertainty due to both
effects was estimated by modifying the missing mass cuts and SIMC model
parameters $p_i$ and investigating the resulting differences on the separated
cross sections.

To estimate the missing mass cut dependence, the experimental and simulated
data were analyzed with two tighter missing mass cuts, $M_X<0.98,\ 1.00$~GeV.
A detailed comparison of the separated cross sections for each $t$-bin
indicated that the $\pi^-$ separated cross sections for higher $-t$ at
\qsq=0.6, 1.0~\gevsq\ \sigl\ were extremely sensitive to the applied $M_X$ cut
and/or the disabling of the collimator pion punch-through routine in the SIMC
simulations.  We believe this is a result of the incomplete $\phi$ coverage for
these settings, as listed in Table \ref{tab:kin}.  The data for any $\pi^-$
$t$-bin were discarded if \sigl\ changed significantly more than the statistical
uncertainty when the nominal $M_X<$1.03 GeV cut is replaced with a $M_X<$1.00
GeV cut in both the experimental and simulation analyses.  For the remaining
$\pi^+$ and $\pi^-$ data, the differences between the ``final'' separated cross
sections and those determined with tighter $M_X$ cuts were computed and the
standard deviation was tabulated for each $-t$ bin at each \qsq.  These
standard deviations for the remaining F$_{\pi}$-1 $\pi^-$ data are in almost
all cases larger than for the corresponding $\pi^+$ data, generally comparable
to the statistical errors.  The standard deviations are typically smallest at
or near $-t_{\text{min}}$ and grow with increasing $-t$.

The cross section model dependence was estimated in a similar manner.  Since
the longitudinal and transverse cross sections in the model reproduce the
experimental values to within 10\%, these two terms were independently
increased and decreased by 10\% in the model.  Independent of this, the
separated cross sections were also determined by alternately setting \siglt=0 and
\sigtt=0 in the model.  Unseparated cross sections were calculated using
Eqn.~\ref{eqn:ratio_to_sigma} and a fit performed using Eqn.~\ref{eqn:unsep}
to extract $L$/$T$/$LT$/$TT$.  The differences between the ``final'' separated
cross sections minus the six independent variations were computed and the
standard deviations tabulated for each $-t$ bin at each \qsq\ in the same
manner as the missing mass cut study.  The model sensitivities of the $L$,$T$
cross sections are generally similar to each other, and exhibit a weaker
$t$-dependence than the $M_X$ cut sensitivities.  The observed variations
are relatively small, about half the statistical uncertainties in
these cross sections (per $t$-bin) of 5-10\%.  The reason is that \sigl\ and
\sigt\ are effectively determined by the $\phi$-integrated cross section, which
reduces the model uncertainty.

The sensitivities of the $TT$ interference response
functions are strongly $t$-dependent, being smaller for the lowest $-t$ bins at
each \qsq\ and increasing for the larger $-t$ bins.  These higher $-t$ bins
have relatively poorer statistics as well as incomplete $\phi$ coverage at low
$\epsilon$ (as well as at high $\epsilon$ for $\pi^-$ \qsq=0.6, 1.0~\gevsq).
The $LT$ model sensitivities are smaller than for $TT$, and show no obvious
trends.

The standard deviations for each \qsq, $t$ bin from the two above studies were
combined in quadrature to obtain the combined systematic uncertainty due to
the missing mass cut and SIMC model dependence (labeled henceforth as
``model-dependence'' for brevity).  The uncertainties computed in this manner are
shown as error bands, presented along with the data in Sec.~\ref{sec:results},
and the values for each bin are listed as the second uncertainty in Tables
\ref{tab:xsec_mn}, \ref{tab:xsec_pl}.

\subsection{Systematic Uncertainties
\label{sec:syst}} 

The various systematic uncertainties determined in Secs.~\ref{sec:expt},
~\ref{sec:analysis} are listed in Tables
\ref{table:Fpi1_syst_unc_pl},~\ref{table:Fpi2_syst_unc_pl}.  Those items not
discussed explicitly in these sections are assumed to be the same as for the
previously published $^1$H analyses.  The systematic uncertainties are
subdivided into correlated and uncorrelated contributions.  The correlated
uncertainties, i.e., those that are the same for both $\epsilon$ points, such
as target thickness corrections, are attributed directly to the separated cross
sections.  Uncorrelated uncertainties are attributed to the unseparated cross
sections, with the result that in the separation of \sigl\ and \sigt\ they are
inflated, just as the statistical uncertainties, by the factor
$1/\Delta\epsilon$ (for \sigl), which is about three.  The uncorrelated
uncertainties can be further subdivided into those that differ in size between
$\epsilon$ points, but may influence the $t$-dependence at a fixed value of
$\epsilon$ in a correlated way.  The largest contributions to the
``$t$-correlated'' uncertainty are acceptance and kinematic offsets, and as a
result, they are the dominating systematic uncertainties for, e.g. \sigl.  In
addition to the uncertainties listed below, are the uncertainties in the
separated cross sections at each $-t$, \qsq\ setting due to the $M_X$ cut and
SIMC model ``model-dependence''.

\section{Results and Discussion
\label{sec:results}}

\subsection{$\bf ^2$H$\bf(e,e'\pi^{\pm})NN_s$ Separated Cross Sections and Ratios}

\begin{figure*}
\begin{center}
\includegraphics[width=0.9\textwidth]{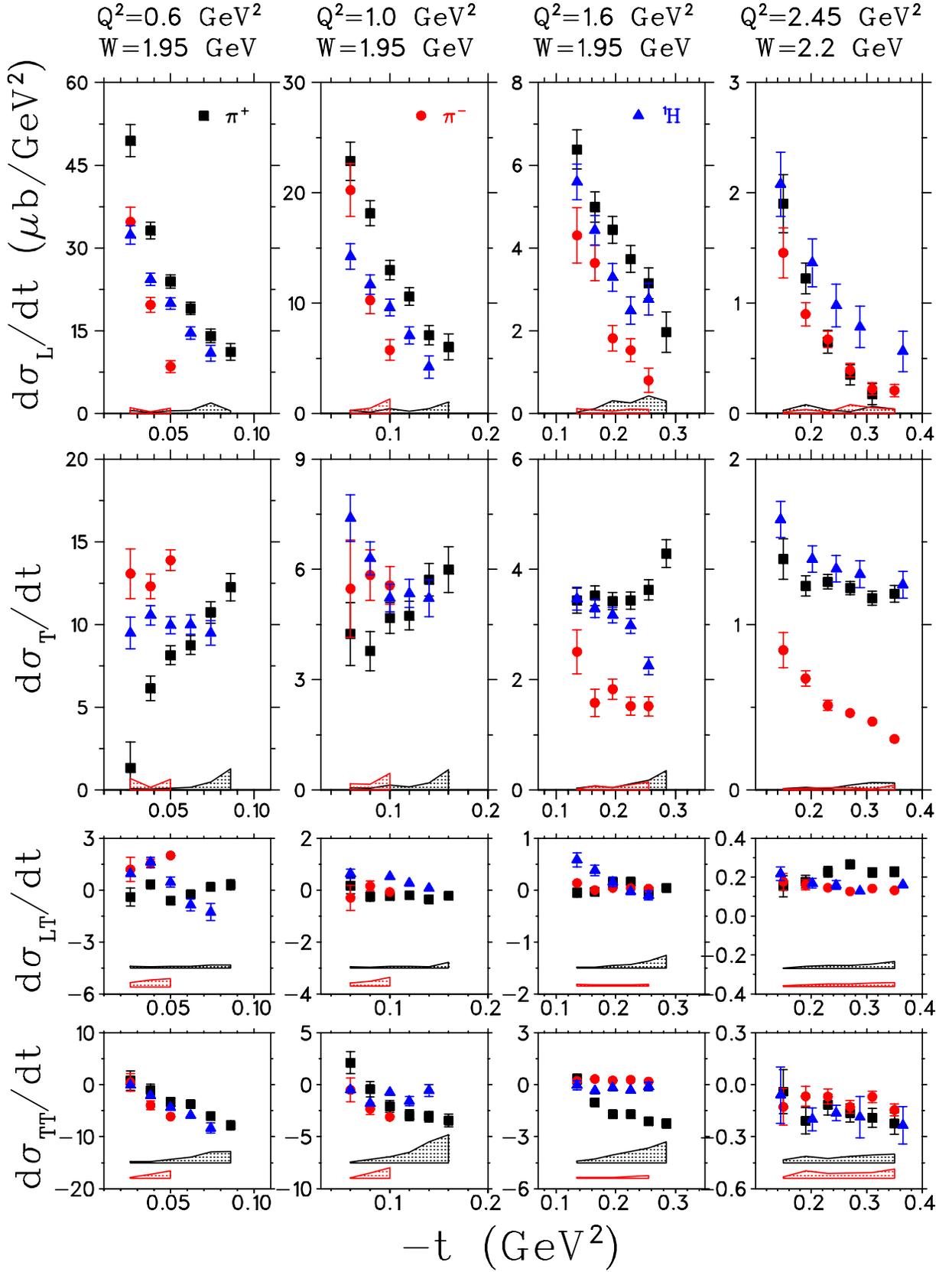}
\caption{\label{fig:xsec} 
(Color online) Separated cross sections as a function of $-t$. 
  {\em $\pi^-$~from~$^2$H:} [red circles], 
  {\em $\pi^+$~from~$^2$H:} [black squares], 
  {\em $\pi^+$~from~$^1$H:} [blue triangles].
  The error bars include both statistical and uncorrelated systematic
  uncertainties.  The ``model-dependences'' of the $L$, $T$, $LT$, $TT$ cross
  sections are indicated by the shaded bands, by which all data points move
  collectively.  The $^1$H data have not been scaled to the mean
  $\overline{Q^2}$, $\overline{W}$ values for each $-t$ bin of $^2$H data.  In
  addition, please keep in mind the issues relating to $^2$H off-shell effects
  before directly comparing the $^1$H and $^2$H data.}
\end{center}
\end{figure*}

\begin{table*}
\begin{center}  
\begin{tabular}{||c|c|c|c|c|c|c||}
\hline  
$\overline{W}$&  $\overline{Q^2}$ & $-t$& $\sigma_T$      & $\sigma_L$      & $\sigma_{TT}$   & $\sigma_{LT}$ \\
 (GeV)        & (GeV$^2$)   & (GeV$^2$) & ($\mu$b/GeV$^2$)& ($\mu$b/GeV$^2$)& ($\mu$b/GeV$^2$)& ($\mu$b/GeV$^2$)\\
\hline\hline
\multicolumn{7}{|c|}{$^2$H$(e,e'\pi^-)pp_s$} \\
\hline\hline
\multicolumn{7}{|c|}{$Q^2=0.60$ GeV$^2$  $W=1.95$ GeV} \\
\hline  
1.9733 & 0.5505 & 0.026 & 13.07 $\pm$ 1.44 $\pm$ 0.69 & 34.74 $\pm$ 2.39 $\pm$ 1.03 &  0.47 $\pm$ 1.17 $\pm$ 0.13 &  1.19 $\pm$ 0.70 $\pm$ 0.26 \\
1.9568 & 0.5765 & 0.038 & 12.31 $\pm$ 0.69 $\pm$ 0.17 & 19.71 $\pm$ 1.13 $\pm$ 0.36 & -3.95 $\pm$ 0.48 $\pm$ 0.74 &  1.54 $\pm$ 0.24 $\pm$ 0.42 \\
1.9452 & 0.6048 & 0.050 & 13.88 $\pm$ 0.62 $\pm$ 0.66 &  8.53 $\pm$ 1.03 $\pm$ 1.01 & -6.13 $\pm$ 0.43 $\pm$ 1.44 &  2.00 $\pm$ 0.20 $\pm$ 0.48 \\
\hline 
\multicolumn{7}{|c|}{$Q^2=1.00$ GeV$^2$  $W=1.95$ GeV} \\
\hline 
1.9864 & 0.9095 & 0.060 &  5.47 $\pm$ 1.29 $\pm$ 0.17 & 20.25 $\pm$ 2.25 $\pm$ 0.29 & -0.50 $\pm$ 0.80 $\pm$ 0.06 & -0.30 $\pm$ 0.48 $\pm$ 0.12 \\
1.9703 & 0.9483 & 0.080 &  5.85 $\pm$ 0.68 $\pm$ 0.16 & 10.27 $\pm$ 1.16 $\pm$ 0.47 & -2.31 $\pm$ 0.38 $\pm$ 0.50 &  0.16 $\pm$ 0.19 $\pm$ 0.19 \\
1.9489 & 0.9977 & 0.100 &  5.56 $\pm$ 0.51 $\pm$ 0.46 &  5.75 $\pm$ 0.91 $\pm$ 1.31 & -3.09 $\pm$ 0.32 $\pm$ 1.01 & -0.08 $\pm$ 0.15 $\pm$ 0.33 \\
\hline 
\multicolumn{7}{|c|}{$Q^2=1.60$ GeV$^2$  $W=1.95$ GeV} \\
\hline  
2.0116 & 1.4345 & 0.135 &  2.51 $\pm$ 0.39 $\pm$ 0.02 & 4.31 $\pm$ 0.66 $\pm$ 0.12 &  0.22 $\pm$ 0.16 $\pm$ 0.04 &  0.13 $\pm$ 0.07 $\pm$ 0.03 \\
1.9867 & 1.5064 & 0.165 &  1.58 $\pm$ 0.24 $\pm$ 0.08 & 3.64 $\pm$ 0.40 $\pm$ 0.09 &  0.33 $\pm$ 0.10 $\pm$ 0.04 & -0.00 $\pm$ 0.05 $\pm$ 0.02 \\
1.9644 & 1.5650 & 0.195 &  1.83 $\pm$ 0.18 $\pm$ 0.05 & 1.82 $\pm$ 0.30 $\pm$ 0.05 &  0.25 $\pm$ 0.08 $\pm$ 0.05 &  0.04 $\pm$ 0.04 $\pm$ 0.03 \\
1.9433 & 1.6178 & 0.225 &  1.52 $\pm$ 0.16 $\pm$ 0.10 & 1.53 $\pm$ 0.27 $\pm$ 0.11 &  0.29 $\pm$ 0.08 $\pm$ 0.09 &  0.04 $\pm$ 0.03 $\pm$ 0.03 \\
1.9229 & 1.6664 & 0.255 &  1.52 $\pm$ 0.18 $\pm$ 0.15 & 0.80 $\pm$ 0.29 $\pm$ 0.10 &  0.19 $\pm$ 0.09 $\pm$ 0.16 &  0.03 $\pm$ 0.03 $\pm$ 0.03 \\
\hline 
\multicolumn{7}{|c|}{$Q^2=2.45$ GeV$^2$  $W=2.22$ GeV} \\
\hline
2.2978 & 2.1619 & 0.150 &  0.85 $\pm$ 0.11 $\pm$ 0.01 & 1.46 $\pm$ 0.22 $\pm$ 0.02 & -0.13 $\pm$ 0.10 $\pm$ 0.01 &  0.18 $\pm$ 0.04 $\pm$ 0.01 \\
2.2695 & 2.2598 & 0.190 &  0.67 $\pm$ 0.05 $\pm$ 0.01 & 0.90 $\pm$ 0.10 $\pm$ 0.04 & -0.07 $\pm$ 0.05 $\pm$ 0.05 &  0.16 $\pm$ 0.03 $\pm$ 0.01 \\
2.2400 & 2.3537 & 0.230 &  0.51 $\pm$ 0.03 $\pm$ 0.02 & 0.67 $\pm$ 0.07 $\pm$ 0.01 & -0.07 $\pm$ 0.04 $\pm$ 0.03 &  0.15 $\pm$ 0.02 $\pm$ 0.02 \\
2.2154 & 2.4289 & 0.270 &  0.47 $\pm$ 0.03 $\pm$ 0.01 & 0.39 $\pm$ 0.06 $\pm$ 0.08 & -0.13 $\pm$ 0.04 $\pm$ 0.03 &  0.13 $\pm$ 0.02 $\pm$ 0.02 \\
2.1932 & 2.4993 & 0.310 &  0.41 $\pm$ 0.02 $\pm$ 0.01 & 0.22 $\pm$ 0.06 $\pm$ 0.06 & -0.07 $\pm$ 0.03 $\pm$ 0.04 &  0.14 $\pm$ 0.02 $\pm$ 0.02 \\
2.1688 & 2.5753 & 0.350 &  0.31 $\pm$ 0.02 $\pm$ 0.03 & 0.21 $\pm$ 0.06 $\pm$ 0.04 & -0.15 $\pm$ 0.03 $\pm$ 0.06 &  0.13 $\pm$ 0.02 $\pm$ 0.02 \\
\hline
\end{tabular}
\end{center}
\caption{\label{tab:xsec_mn} Separated cross sections for the
  $^2$H$(e,e'\pi^-)pp_s$ reaction.  The first uncertainties listed are
  statistical only.  The second uncertainties listed are the $M_X$ cut and SIMC
  model ``model-dependences''.  In addition to these, the systematic
  uncertainties listed in Tables \ref{table:Fpi1_syst_unc_pl} and
  \ref{table:Fpi2_syst_unc_pl} must be applied.}
\end{table*}

\begin{table*}
\begin{center}  
\begin{tabular}{||c|c|c|c|c|c|c||}
\hline  
$\overline{W}$&  $\overline{Q^2}$ & $-t$& $\sigma_T$      & $\sigma_L$      & $\sigma_{TT}$   & $\sigma_{LT}$ \\
 (GeV)        & (GeV$^2$)   & (GeV$^2$) & ($\mu$b/GeV$^2$)& ($\mu$b/GeV$^2$)& ($\mu$b/GeV$^2$)&($\mu$b/GeV$^2$)\\
\hline\hline
\multicolumn{7}{|c|}{$^2$H$(e,e'\pi^+)nn_s$} \\
\hline\hline
\multicolumn{7}{|c|}{$Q^2=0.60$ GeV$^2$  $W=1.95$ GeV} \\
\hline 
1.9702 & 0.5445 & 0.026 &  1.32 $\pm$ 1.49 $\pm$ 0.10 & 49.44 $\pm$ 2.51 $\pm$ 0.56 &  0.80 $\pm$ 1.11 $\pm$ 0.21 & -0.40 $\pm$ 0.53 $\pm$ 0.08 \\
1.9572 & 0.5736 & 0.038 &  6.15 $\pm$ 0.64 $\pm$ 0.06 & 33.17 $\pm$ 1.18 $\pm$ 0.16 & -1.06 $\pm$ 0.56 $\pm$ 0.24 &  0.32 $\pm$ 0.26 $\pm$ 0.07 \\ 
1.9495 & 0.5953 & 0.050 &  8.15 $\pm$ 0.51 $\pm$ 0.12 & 23.94 $\pm$ 0.97 $\pm$ 0.47 & -3.33 $\pm$ 0.46 $\pm$ 0.65 & -0.61 $\pm$ 0.20 $\pm$ 0.10 \\ 
1.9444 & 0.6092 & 0.062 &  8.76 $\pm$ 0.54 $\pm$ 0.17 & 19.08 $\pm$ 0.99 $\pm$ 0.54 & -3.73 $\pm$ 0.49 $\pm$ 1.02 & -0.25 $\pm$ 0.21 $\pm$ 0.11 \\
1.9423 & 0.6146 & 0.074 & 10.73 $\pm$ 0.64 $\pm$ 0.48 & 14.08 $\pm$ 1.15 $\pm$ 1.90 & -5.99 $\pm$ 0.61 $\pm$ 2.04 &  0.19 $\pm$ 0.23 $\pm$ 0.17 \\
1.9411 & 0.6206 & 0.086 & 12.25 $\pm$ 0.81 $\pm$ 1.29 & 11.18 $\pm$ 1.45 $\pm$ 0.53 & -7.84 $\pm$ 0.83 $\pm$ 2.19 &  0.30 $\pm$ 0.29 $\pm$ 0.18 \\
\hline  
\multicolumn{7}{|c|}{$Q^2=0.75$ GeV$^2$  $W=1.95$ GeV} \\
\hline
1.9894 & 0.6668 & 0.037 &  8.76 $\pm$ 1.22 $\pm$ 0.15 & 21.76 $\pm$ 2.03 $\pm$ 0.48 &  2.13 $\pm$ 0.68 $\pm$ 0.18 &  0.67 $\pm$ 0.29 $\pm$ 0.02 \\ 
1.9691 & 0.6978 & 0.051 & 10.82 $\pm$ 0.80 $\pm$ 0.29 & 15.90 $\pm$ 1.32 $\pm$ 0.39 & -0.54 $\pm$ 0.42 $\pm$ 0.38 &  0.42 $\pm$ 0.18 $\pm$ 0.07 \\
1.9579 & 0.7259 & 0.065 & 10.34 $\pm$ 0.66 $\pm$ 0.34 & 14.41 $\pm$ 1.11 $\pm$ 0.28 & -3.70 $\pm$ 0.38 $\pm$ 0.75 &  0.54 $\pm$ 0.15 $\pm$ 0.10 \\
1.9467 & 0.7483 & 0.079 &  9.36 $\pm$ 0.64 $\pm$ 0.29 & 16.06 $\pm$ 1.08 $\pm$ 1.65 & -6.93 $\pm$ 0.42 $\pm$ 1.42 &  0.22 $\pm$ 0.13 $\pm$ 0.11 \\
1.9404 & 0.7640 & 0.093 &  9.75 $\pm$ 0.69 $\pm$ 0.37 & 15.82 $\pm$ 1.18 $\pm$ 4.73 & -9.57 $\pm$ 0.52 $\pm$ 2.41 &  0.39 $\pm$ 0.15 $\pm$ 0.23 \\
1.9357 & 0.7805 & 0.107 & 11.10 $\pm$ 0.81 $\pm$ 0.58 & 13.76 $\pm$ 1.38 $\pm$ 7.22 &-12.50 $\pm$ 0.69 $\pm$ 3.45 &  1.12 $\pm$ 0.16 $\pm$ 0.41 \\
\hline 
\multicolumn{7}{|c|}{$Q^2=1.00$ GeV$^2$  $W=1.95$ GeV} \\
\hline 
1.9970 & 0.8941 & 0.060 & 4.24 $\pm$ 0.82 $\pm$ 0.06 & 22.87 $\pm$ 1.55 $\pm$ 0.28 &  2.13 $\pm$ 0.71 $\pm$ 0.08 &  0.17 $\pm$ 0.31 $\pm$ 0.04 \\
1.9802 & 0.9305 & 0.080 & 3.78 $\pm$ 0.50 $\pm$ 0.05 & 18.16 $\pm$ 0.95 $\pm$ 0.12 & -0.42 $\pm$ 0.41 $\pm$ 0.31 & -0.25 $\pm$ 0.18 $\pm$ 0.04 \\
1.9602 & 0.9745 & 0.100 & 4.68 $\pm$ 0.40 $\pm$ 0.14 & 13.00 $\pm$ 0.76 $\pm$ 0.45 & -2.07 $\pm$ 0.35 $\pm$ 0.59 & -0.23 $\pm$ 0.13 $\pm$ 0.06 \\
1.9458 & 1.0061 & 0.120 & 4.74 $\pm$ 0.37 $\pm$ 0.09 & 10.60 $\pm$ 0.72 $\pm$ 0.20 & -2.93 $\pm$ 0.36 $\pm$ 1.01 & -0.20 $\pm$ 0.12 $\pm$ 0.06 \\
1.9349 & 1.0320 & 0.140 & 5.72 $\pm$ 0.44 $\pm$ 0.20 &  7.10 $\pm$ 0.83 $\pm$ 0.45 & -3.07 $\pm$ 0.43 $\pm$ 2.03 & -0.36 $\pm$ 0.13 $\pm$ 0.05 \\
1.9247 & 1.0602 & 0.160 & 6.00 $\pm$ 0.62 $\pm$ 0.55 &  6.04 $\pm$ 1.14 $\pm$ 1.05 & -3.44 $\pm$ 0.58 $\pm$ 2.69 & -0.22 $\pm$ 0.16 $\pm$ 0.21 \\
\hline 
\multicolumn{7}{|c|}{$Q^2=1.60$ GeV$^2$  $W=1.95$ GeV} \\
\hline  
2.0112 & 1.4353 & 0.135 & 3.43 $\pm$ 0.22 $\pm$ 0.03 &  6.38 $\pm$ 0.43 $\pm$ 0.03 &  0.34 $\pm$ 0.22 $\pm$ 0.09 & -0.05 $\pm$ 0.09 $\pm$ 0.01 \\
1.9884 & 1.4998 & 0.165 & 3.52 $\pm$ 0.17 $\pm$ 0.07 &  5.00 $\pm$ 0.34 $\pm$ 0.12 & -1.01 $\pm$ 0.16 $\pm$ 0.23 & -0.03 $\pm$ 0.06 $\pm$ 0.02 \\
1.9669 & 1.5553 & 0.195 & 3.43 $\pm$ 0.15 $\pm$ 0.05 &  4.44 $\pm$ 0.30 $\pm$ 0.31 & -1.70 $\pm$ 0.16 $\pm$ 0.44 &  0.16 $\pm$ 0.05 $\pm$ 0.05 \\
1.9463 & 1.6082 & 0.225 & 3.44 $\pm$ 0.15 $\pm$ 0.12 &  3.74 $\pm$ 0.30 $\pm$ 0.26 & -1.70 $\pm$ 0.17 $\pm$ 0.65 &  0.16 $\pm$ 0.05 $\pm$ 0.06 \\
1.9276 & 1.6568 & 0.255 & 3.63 $\pm$ 0.18 $\pm$ 0.18 &  3.15 $\pm$ 0.36 $\pm$ 0.43 & -2.10 $\pm$ 0.20 $\pm$ 0.86 & -0.02 $\pm$ 0.05 $\pm$ 0.13 \\
1.9097 & 1.7025 & 0.285 & 4.29 $\pm$ 0.25 $\pm$ 0.36 &  1.97 $\pm$ 0.48 $\pm$ 0.30 & -2.24 $\pm$ 0.26 $\pm$ 1.20 &  0.04 $\pm$ 0.07 $\pm$ 0.25 \\
\hline 
\multicolumn{7}{|c|}{$Q^2=2.45$ GeV$^2$  $W=2.22$ GeV} \\
\hline
2.3017 & 2.1503 & 0.150 & 1.40 $\pm$ 0.12 $\pm$ 0.01 & 1.90 $\pm$ 0.26 $\pm$ 0.03 & -0.04 $\pm$ 0.12 $\pm$ 0.02 &  0.15 $\pm$ 0.05 $\pm$ 0.01 \\
2.2719 & 2.2518 & 0.190 & 1.23 $\pm$ 0.06 $\pm$ 0.02 & 1.22 $\pm$ 0.14 $\pm$ 0.08 & -0.21 $\pm$ 0.07 $\pm$ 0.04 &  0.18 $\pm$ 0.03 $\pm$ 0.01 \\
2.2448 & 2.3391 & 0.230 & 1.26 $\pm$ 0.04 $\pm$ 0.01 & 0.65 $\pm$ 0.11 $\pm$ 0.03 & -0.12 $\pm$ 0.06 $\pm$ 0.03 &  0.23 $\pm$ 0.03 $\pm$ 0.02 \\
2.2197 & 2.4180 & 0.270 & 1.22 $\pm$ 0.04 $\pm$ 0.03 & 0.35 $\pm$ 0.09 $\pm$ 0.02 & -0.16 $\pm$ 0.05 $\pm$ 0.04 &  0.26 $\pm$ 0.02 $\pm$ 0.02 \\
2.1977 & 2.4878 & 0.310 & 1.16 $\pm$ 0.04 $\pm$ 0.05 & 0.18 $\pm$ 0.10 $\pm$ 0.06 & -0.19 $\pm$ 0.06 $\pm$ 0.05 &  0.22 $\pm$ 0.02 $\pm$ 0.02 \\
2.1750 & 2.5570 & 0.350 & 1.19 $\pm$ 0.05 $\pm$ 0.04 &-0.10 $\pm$ 0.11 $\pm$ 0.04 & -0.22 $\pm$ 0.06 $\pm$ 0.05 &  0.23 $\pm$ 0.02 $\pm$ 0.04 \\
\hline
\end{tabular}
\end{center}
\caption{\label{tab:xsec_pl} Separated cross sections for the
  $^2$H$(e,e'\pi^+)nn_s$ reaction.  The first uncertainties listed are
  statistical only.  The second uncertainties listed are the $M_X$ cut and SIMC
  model ``model-dependences''.  In addition to these, the systematic
  uncertainties listed in Tables \ref{table:Fpi1_syst_unc_pl} and
  \ref{table:Fpi2_syst_unc_pl} must be applied.}
\end{table*}

\begin{figure*}
\begin{center}
\includegraphics[angle=90,width=6.5in]{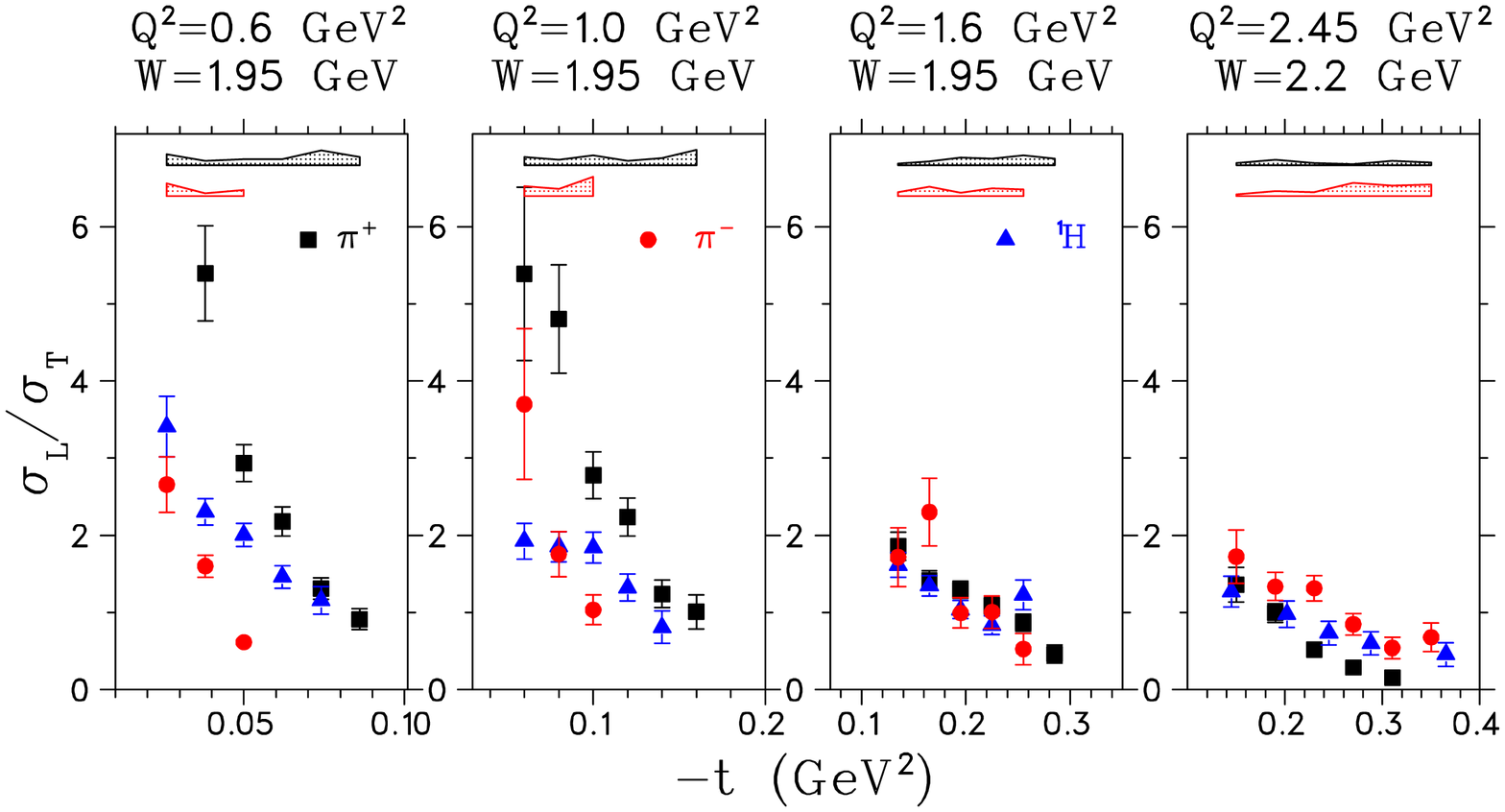}
\caption{\label{fig:ltratio}
(Color online) $L$/$T$ separated cross section ratios as a function of $-t$ for
  $\pi^+$ [black squares] and $\pi^-$ [red circles] production on $^2$H, and
  for $\pi^+$ on $^1$H [blue triangles].  The model-dependences of the ratios
  are indicated by the shaded bands, by which all data points move
  collectively.}
\end{center}
\end{figure*}

The $\pi^{\pm}$ separated cross sections from $^2$H are shown in
Fig.~\ref{fig:xsec} and are listed in Tables~\ref{tab:xsec_mn} and
\ref{tab:xsec_pl}.  Also shown for comparison are our previously published
$\pi^+$ data from $^1$H \cite{Blok08}.  Please keep in mind the issues relating
to $^2$H off-shell effects discussed in Sec.~\ref{sec:model} before directly
comparing the $^1$H and $^2$H data, particularly at higher $-t$, where the
effect of Fermi momentum is larger.

In the $L$ response of Fig. \ref{fig:xsec}, the pion pole is evident by the sharp
rise at small $-t$.  The cross sections for $\pi^-$ and $\pi^+$ from $^2$H are
similar to each other and to those from $^1$H, but there is a general tendency
for the $\pi^-$ \sigl\ to drop more rapidly with $-t$ than the $\pi^+$ \sigl.

The $T$ responses are much flatter versus $t$.  With the exception of the lowest
two $-t$ bins at \qsq=0.6 \gevsq, the $\pi^+$ \sigt\ from $^2$H are generally
within the uncertainties of the \sigt\ from $^1$H.  We have looked very
carefully at the analysis of these two low $-t$ bins, but we were
unable to identify a specific reason for this behavior, hence we do not
believe it is due to an artifact of the analysis.  We note that these two
$-t$ bins correspond to the smallest relative momentum of the two recoil
nucleons in our data set ($<$170 MeV/c), where nucleonic final-state
interactions absent for $^1$H may be relevant.

It is also seen that the $\pi^-$ \sigt\ are
significantly lower than the $\pi^+$ \sigt\ at \qsq=1.6, 2.45 \gevsq.  The
suppression of $\sigma_T^{\pi^-}$ relative to $\sigma_T^{\pi^+}$ may benefit
future measurements of $F_{\pi}(Q^2)$ since the larger $L$/$T$ ratio in
$^2$H$(e,e'\pi^-)pp_s$ would enjoy reduced error magnification compared to
$p(e,e'\pi^+)nn_s$.  This enhancement in the $L$/$T$ ratio at higher \qsq\ is
seen more clearly in Fig.  \ref{fig:ltratio}.

The interference \siglt, \sigtt\ cross sections are shown in the bottom two
rows of Fig.~\ref{fig:xsec}.  Interestingly, at higher \qsq\ the $\pi^-$
interference cross sections are more similar to the $\pi^+$ cross sections from
$^1$H than from $^2$H.  Also note that the model-dependence of the interference
cross sections grows dramatically with $-t$, particularly for the $\pi^+$ cross
sections from $^2$H.  The model-dependences from $^1$H are not shown, but are
significantly smaller.

\begin{figure}
\begin{center}
\includegraphics[height=3.25in]{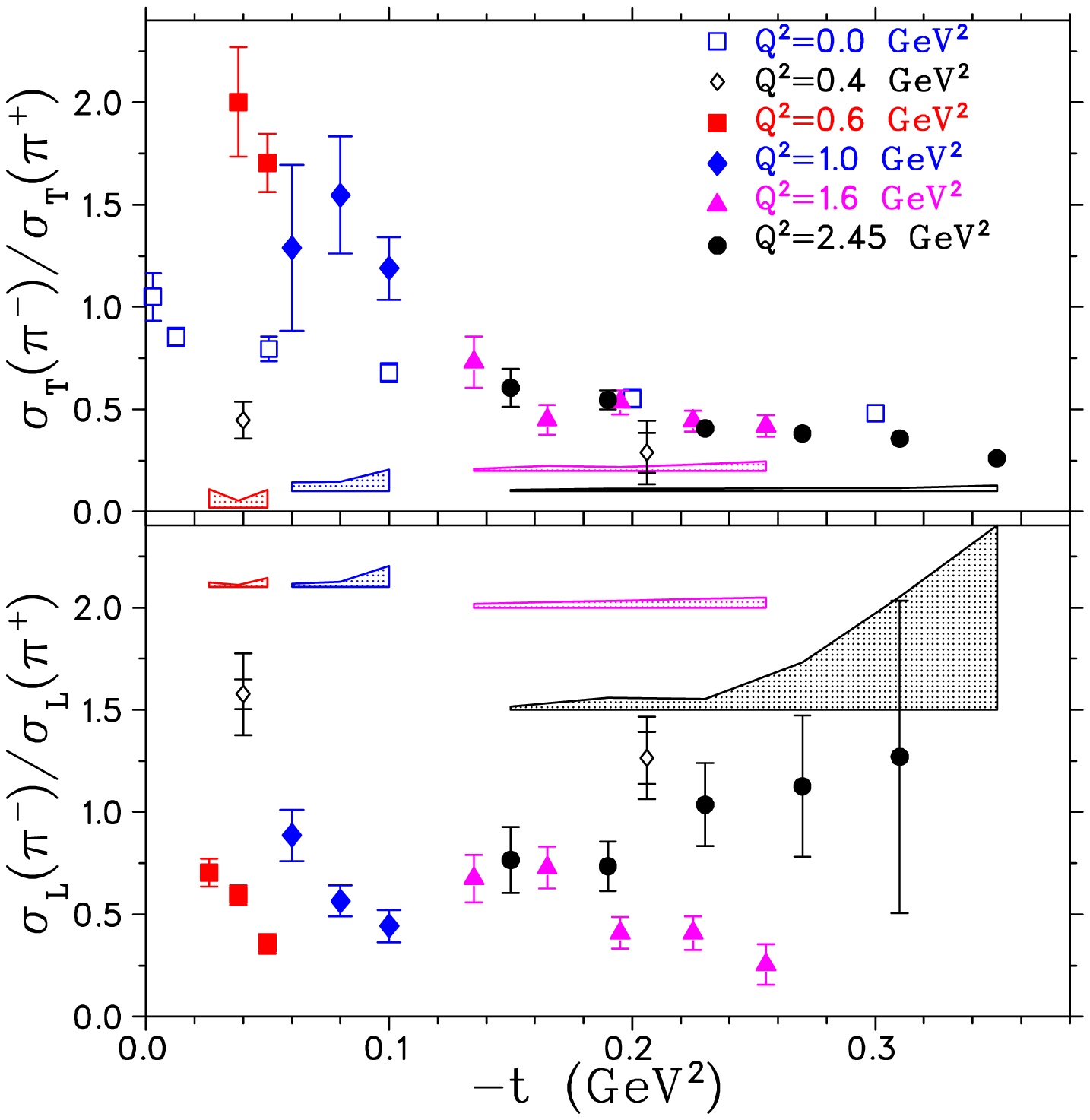}
\end{center}
\caption{(Color online) The ratios $R_L$ and $R_T$ versus $-t$ for four
  \qsq\ settings from this work.  The model-dependences of the ratios are
  indicated by the shaded bands, and the error bars include statistical and
  uncorrelated systematic uncertainties.  Also shown are the ratios at \qsq=0.4
  \gevsq\ in the resonance region from Refs. \cite{Gaskellthesis, Gaskell01},
  and $R_T$ from the $E_{\gamma}$=3.4 GeV photoproduction data of
  Ref.~\cite{heide}.
  \label{fig:Rlt_plot1}}
\end{figure}

$\pi^-/\pi^+$ ratios of the separated cross sections were formed, in which
nuclear binding and rescattering effects are expected to cancel.
(No corrections have been made for electromagnetic FSI or two-photon exchange
effects, but these are expected to be small.)  Many experimental normalization
factors cancel to a high degree in the ratio (acceptance, target thickness,
pion decay and absorption in the detectors, radiative corrections, etc.).  The
principal remaining uncorrelated systematic errors are in the tracking
inefficiencies, target boiling corrections (due to different beam currents
used), and \u{C}erenkov blocking corrections.

Figure \ref{fig:Rlt_plot1} shows the values of the separated cross section
$\pi^-/\pi^+$ ratios.  $R_L$ is approximately 0.8 near $-t_{\text{min}}$ at each
\qsq\ setting, as predicted in the large $N_c$ limit calculation of
Ref.~\cite{frankfurt}.  Under the not necessarily realistic assumption that the
isoscalar and isovector amplitudes are real, $R_L=0.8$ gives $A_S/A_V=6\%$.
This is relevant for the extraction of the pion form factor from
electroproduction data, which uses a model including some isoscalar background.
It is difficult at this stage to make a more quantitative conclusion, but this
result is qualitatively in agreement with the findings of our pion form factor
analyses \cite{Huber08,volmer}, which found evidence of a small additional
contribution to \sigl\ not taken into account by the VGL Regge Model in our
\qsq=0.6-1.6~\gevsq\ data at $W=1.95$ GeV, but little evidence for any
additional contributions in our \qsq=1.6-2.45 \gevsq\ data at $W=2.2$ GeV.  The
main conclusion to be drawn is that pion exchange dominates the forward
longitudinal response even $\sim 10\ m_{\pi}^2$ away from the pion pole.
The $R_L$ results from Gaskell, et al. \cite{Gaskellthesis, Gaskell01} at
\qsq=0.4~\gevsq, $W<1.7$ GeV, are above 1, presumably because of significant
resonance contributions.

Also in Fig.~\ref{fig:Rlt_plot1} are our $R_T$ results, following a nearly
universal curve with $-t$, and exhibiting only a small \qsq-dependence.
Interestingly, above $-t=0.15$ \gevsq, the photoproduction $R_T$ at
$E_{\gamma}$=3.4 GeV from Heide, et al., \cite{heide} are very close in value
to our ratios from electroproduction.  Of the \qsq=0.4~\gevsq\ data from
Refs. \cite{Gaskellthesis, Gaskell01}, the higher $-t$ point [$-t=0.21$
\gevsq\ at $W=1.15$ GeV, below the $\Delta_{1232}$] is closer to the
`universal curve', while the lower $-t$ point [$-t=0.04$ \gevsq\ at $W=1.6$
  GeV, in the resonance region] is well below it.

At the highest \qsq\ and $-t$, $R_T$ reaches $0.26\pm 0.02$, which is
consistent with the $s$-channel knockout of valence quarks prediction by
Nachtmann \cite{nachtmann},
\begin{equation}
\frac{\gamma^*_T n\rightarrow\pi^-p}{\gamma^*_T
  p\rightarrow\pi^+n}=\Bigl(\frac{e_d}{e_u}\Bigr)^2=\frac{1}{4},
\end{equation}
at sufficiently large $-t$.  This value is reached at a much lower value of
$-t$ than for the unseparated ratios of Ref.~\cite{Brauel1}.
A value of $-t=0.3$ \gevsq\ seems quite a low value for quark charge scaling
arguments to apply directly.  This might indicate the partial cancellation of
soft QCD corrections in the formation of the $\pi^-/\pi^+$ ratio.  Data at
larger $-t$ are needed to see if this interpretation is correct.

Photoproduction data \cite{Gao} at $-t\geq3$ \gevsq\ have hinted at
quark-partonic behavior, based on the combination of constituent scaling, and
experimental results for $R_T$.  However, the experimental photoproduction
cross sections are much larger than can be accounted for by
one-hard-gluon-exchange diagrams in a handbag factorization calculation, even
at $s\sim 10$ \gevsq\ \cite{huang00}.  Either the vector meson dominance
contribution is still large, or the leading-twist generation of the meson
underestimates the handbag contribution \cite{kroll02}.  However, by forming
the $\pi^-/\pi^+$ ratio the nonperturbative components represented by the form
factors and meson distribution amplitude may be divided out, allowing the
perturbative contribution to be observed more readily.  In the limit that the
soft contributions are completely divided out, the one-hard-gluon-exchange
calculation predicts \cite{kroll02} the simple scaling behavior
\begin{displaymath}
\frac{d\sigma(\gamma n\rightarrow\pi^- p)}{d\sigma (\gamma p\rightarrow
\pi^+n)} \approx \Bigl[ \frac{e_d (u-m_p^2)+ e_u(s-m_p^2)}{e_u (u-m_p^2)+
e_d(s-m_p^2)} \Bigr]^2.
\end{displaymath}
The recent JLab data at $\theta_{CM}=90^o$ and above $-t=3$ \gevsq\ are in
agreement with the above expression, while those at smaller $\theta_{CM}$ are not
\cite{Gao}.

A possible explanation for the relatively early perturbative behavior in
transverse electroproduction is that the quasi-free process $e q\rightarrow e
q$ has the minimal total number of elementary fields (4) \cite{brodsky73} and
so requires only a single photon exchange.  The fact that only a single pion is
created may be crucial to this quasi-free picture, since it implies that the
string tension never greatly exceeds O($m_{\pi}$).  By contrast, the
photoproduction reaction $\gamma q\rightarrow q$ at high $-t$ can only proceed
if the initial quark is far off its mass shell. The required strong binding to
other quarks leads to the larger number of active elementary fields in $\gamma
N \rightarrow \pi N$ (9) and hence $s^{2-n}=s^{-7}$ scaling.

\begin{figure*}
\begin{center}
\includegraphics[angle=90,width=6.5in]{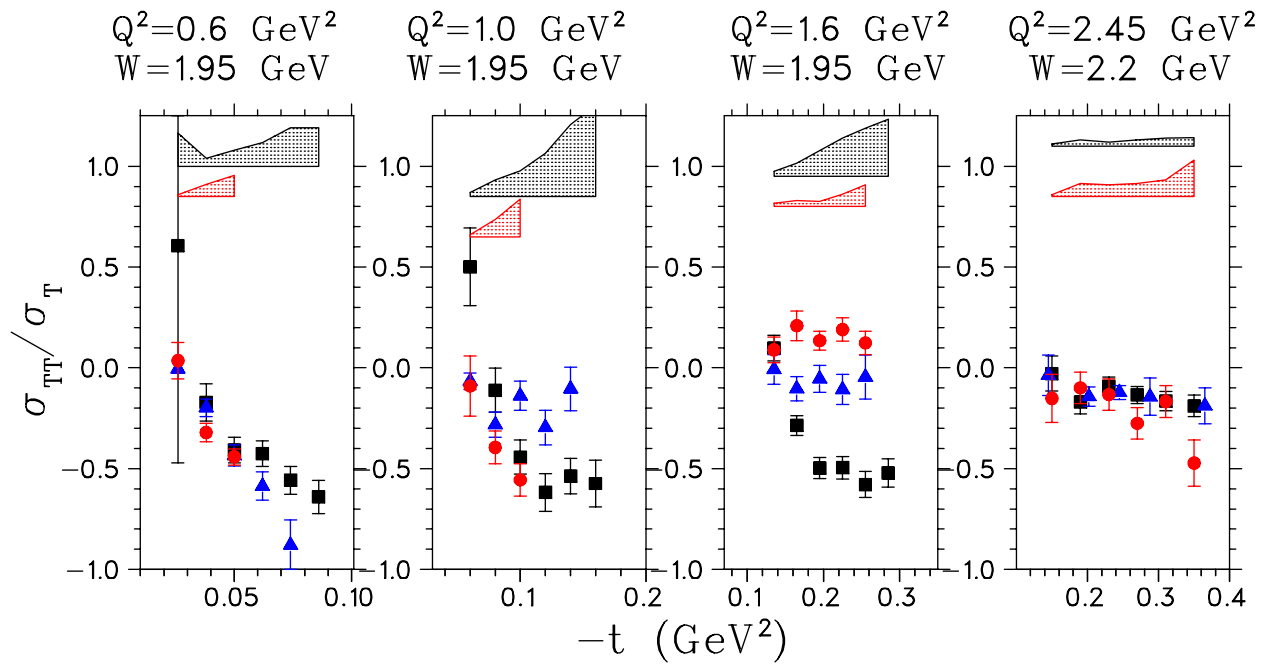}
\caption{\label{fig:tttratio}
(Color online) $TT$/$T$ separated cross section ratios as a function of $-t$.
  The legend is the same as in Fig.~\protect{\ref{fig:ltratio}}.}
\end{center}
\end{figure*}

Another prediction of the quark-parton mechanism \cite{nachtmann} is the
suppression of $\sigma_{TT}/\sigma_T$ due to $s$-channel helicity conservation.
Our data support this hypothesis in that \sigtt\ decreases more rapidly than
\sigt\ with increasing \qsq.  This is particularly true for $\pi^+$
electroproduction on both $^2$H and $^1$H, where $\sigma_{TT}/\sigma_T\simeq
(-19\pm 1)\%$ at our highest \qsq, $-t$ setting (see Fig.~ \ref{fig:tttratio}).
The $\sigma_{TT}/\sigma_T$ ratios for $\pi^-$ production are generally
consistent with those for $\pi^+$, once one takes into account the respective
error bars and model-dependences.

\subsection{Comparison of Various Models with the Data}

\begin{figure*}
\begin{center}
\includegraphics[width=0.9\textwidth]{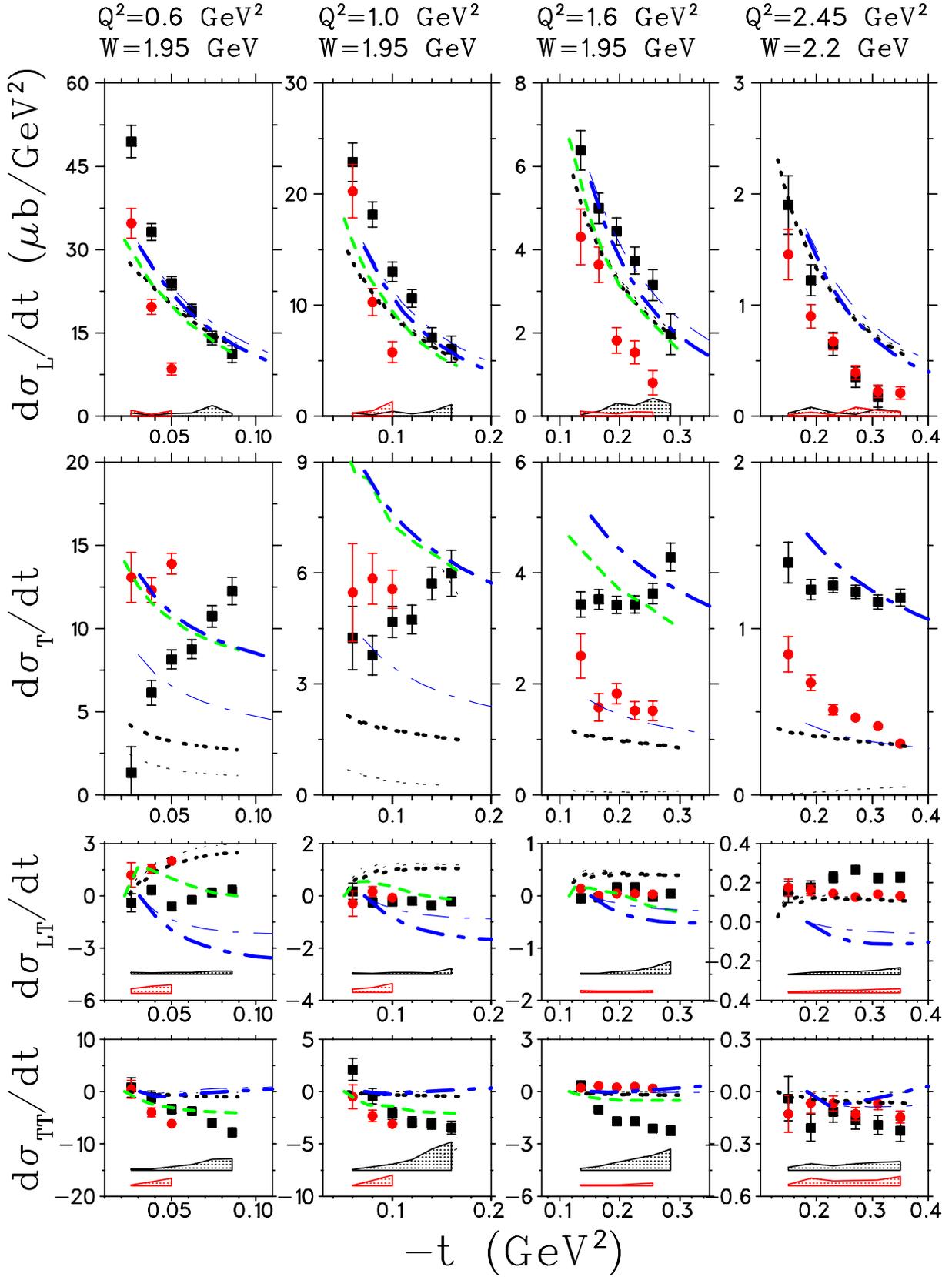}
\caption{\label{fig:xsec_models1} 
(Color online) Comparison of separated cross sections as a function of $-t$
  with various models.  {\em $\pi^-$~from~$^2$H:} [red circles], {\em
  $\pi^+$~from~$^2$H:}[black squares].  The data error bars and
  bands are as in Fig.~\ref{fig:xsec}.  The dotted black curves are predictions
  of the VGL Regge model \protect{\cite{VGL1}} using the values
  $\Lambda_{\pi}^2$=0.0.394, 0.411, 0.455, 0.491~\gevsq\ and
  $\Lambda_{\rho}$=1.50~\gevsq, as determined from fits to our $^1$H data
  \protect{\cite{Huber08}}.  The short-dashed green curves are predictions by
  Kaskulov and Mosel \protect{\cite{kaskulov}}, and the dot-dashed blue curves
  are the predictions by Vrancx and Ryckebusch \protect{\cite{vrancx}}, both
  models are evaluated at the nominal kinematics.  In all cases, the thick
  lines are the model predictions for $\pi^+$ and the thin lines are the
  predictions for $\pi^-$.}
\end{center}
\end{figure*}

\begin{figure*}
\begin{center}
\includegraphics[width=0.9\textwidth]{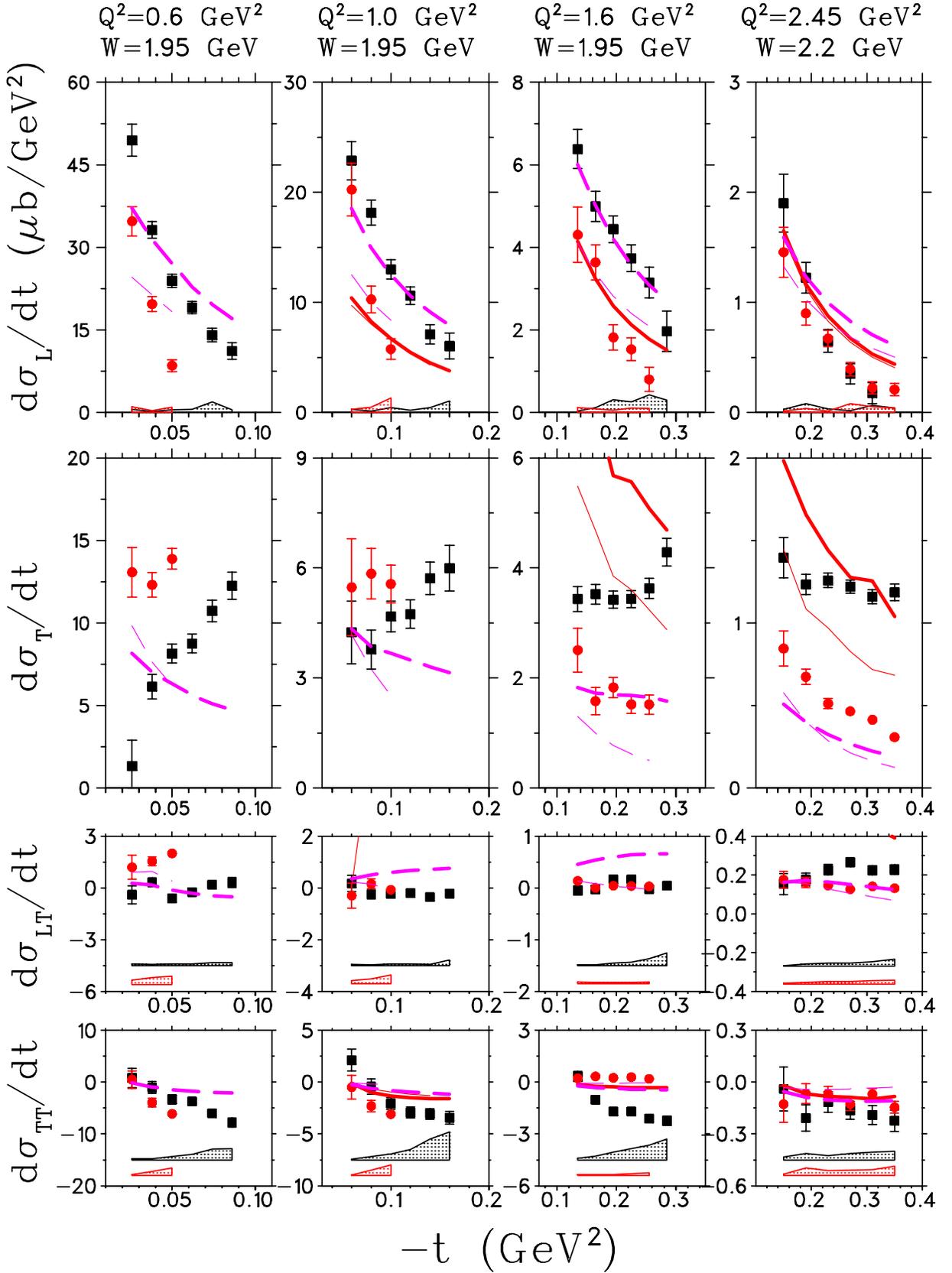}
\caption{\label{fig:xsec_models2} 
(Color online) Comparison of separated cross sections as a function of $-t$
  with various models.  The symbols are as in Fig.~\ref{fig:xsec_models1}.  The
  long-dashed magenta curves are the predictions of the MAID07 model
  \protect{\cite{maid07}}, and the solid red curves are predictions by
  Goloskokov and Kroll \protect{\cite{gk13}}.  Both models are evaluated at
  the same $\overline{W}$, $\overline{Q^2}$ as the data.  The thick lines are
  the model predictions for $\pi^+$ and the thin lines are the predictions for
  $\pi^-$.}
\end{center}
\end{figure*}

\begin{figure*}
\begin{center}
\includegraphics[height=6.5in, angle=90.]{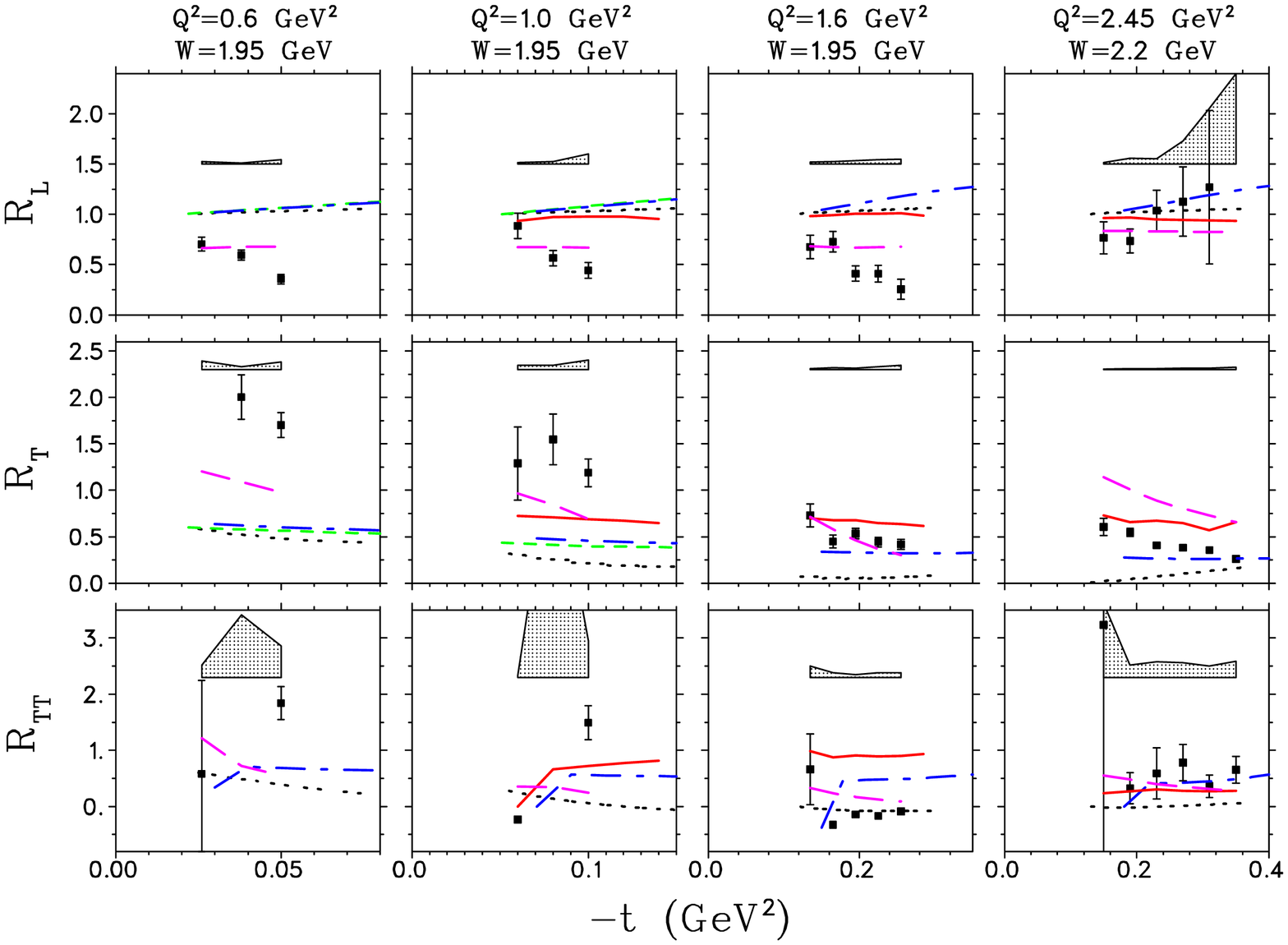}
\end{center}
\caption{(Color online) The ratios $R_L$, $R_T$ and
$R_{TT}\equiv\sigma_{TT}^{\pi^-}/\sigma_{TT}^{\pi^+}$ versus $-t$ for four
\qsq\ settings.  The error bars include statistical and uncorrelated systematic
uncertainties.  The model-dependences of the ratios are indicated by the shaded
bands.  The model legend is the same as Figs. \ref{fig:xsec_models1},
\ref{fig:xsec_models2}; i.e. dotted black curves are the VGL Regge model
\protect{\cite{VGL1}}, short-dashed green curves are Kaskulov and Mosel
\protect{\cite{kaskulov}}, dot-dashed blue curves are Vrancx and Ryckebusch
\protect{\cite{vrancx}}, long-dashed magenta curves are the MAID07 model
\protect{\cite{maid07}}, and solid red curves are Goloskokov and Kroll
\protect{\cite{gk13}}.
\label{fig:Rlt_plot}}
\end{figure*}

The separated cross section data are compared to a variety of models in
Figs. \ref{fig:xsec_models1},~\ref{fig:xsec_models2}, and our $R_L$, $R_T$ and
$R_{TT}\equiv\sigma_{TT}^{\pi^-}/\sigma_{TT}^{\pi^+}$ ratios are compared to
the same models in Fig.~\ref{fig:Rlt_plot}.

The VGL Regge model, which does well for photoproduction \cite{VGL1} and
longitudinal electroproduction \cite{Blok08}, fails to describe the magnitude
or the \qsq-dependence of \sigt.  For any choice for
the $\rho \pi \gamma$ monopole mass, it underpredicts the transverse cross
sections by a large factor, which increases with \qsq.  As briefly
mentioned in the introduction, the VGL Regge model was extended by Kaskulov and
Mosel (KM) \cite{kaskulov} and more recently by Vrancx and Ryckebusch (VR)
\cite{vrancx}.  KM add to the Regge model a hadronic model, which incorporates
DIS $\pi^{\pm}$ electroproduction at the amplitude level.  This DIS process
dominates the transverse response at moderate and high \qsq, increasing the
predicted \sigt.  In this approach, the residual effects of nucleon resonances
in the proton electromagnetic transition form factor are treated as dual to
partons, i.e. ``resonance-parton (R-P) contributions''.  The VR model differs
from the KM model by using an alternative R-P transition form factor, which
better describes the deep-inelastic $N(e,e'\pi^{\pm})$ data.

The VGL model parameters used here are taken from the fits to our $^1$H
\sigl\ data shown in Ref.~\cite{Huber08}.  Similarly, the KM and VR models base
their parameterization of the pion electromagnetic form factor upon fits to our
$^1$H \sigl\ data.  Not surprisingly, the VGL and KM models predict nearly
identical \sigl\ in Fig.~\ref{fig:xsec_models1}, while the VR values are a bit
higher.  For \sigt, the KM and VR models are much closer to the experimental
values than VGL, but they predict a steeper $t$-dependence than exhibited by
the data.  Of these three models, KM also provides the best description of the
$\pi^+$ \siglt\ and \sigtt\ data.

The $R_L$ predictions of the VGL, KM and VR models are nearly identical at
\qsq=0.6, 1.0 \gevsq, with some differences becoming apparent at larger
\qsq\ and $-t$.  With the exception of the highest $-t$ points at \qsq=2.45
\gevsq, the models generally predict $R_L$ ratios that are too large in
comparison to the data.  As already discussed, the reason for this discrepancy
for the three \qsq\ taken at $W=1.95$ GeV is believed to be a small resonance
contribution in the longitudinal channel that is not included in these models.
The VGL, KM and VR models also generally underpredict $R_T$, particularly at
$-t_{\text{min}}$.  However, the KM and VR models predict systematically larger
$R_T$ values than VGL due to the addition of the DIS mechanism to the
transverse channel.  In fact, the VR model comes quite close to the data at
higher $-t$, and \qsq, validating their improvements to the R-P transition form
factor, such as a softer proton Dirac form factor.


The MAID model is a phenomenological fit to pion electroproduction data in the
canonical resonance region ($W<2$ GeV). This model incorporates Breit-Wigner
fits to nucleon resonances and also includes (unitarized) non-resonant
backgrounds.  Originally introduced in 1998 \cite{maid98}, MAID has undergone
incremental improvements.  Shown here are the results of the most recent
version of the MAID model from 2007 \cite{maid07}.  For these calculations, we
have used the MAID07 standard parameter set, although some parameters (such as
relative strengths of resonances, the charged pion form factor, etc.) can be
adjusted.  Finally, note that we apply the MAID model to some kinematics with
$W>2$ GeV. Strictly speaking, the model is not constrained in this regime and
the results plotted represent an extrapolation of a calculation fit at lower
$W$.

For \sigl, the MAID07 predictions are slightly higher than the VGL, KM and VR
models, while the \sigt\ predictions are midway between the purely Regge-based
VGL and the VGL+DIS KM and VR.  In terms of $\pi^-/\pi^+$ ratios, MAID07
provides by far the best description of $R_L$, providing further evidence
that the disagreement between the pion-pole dominated models and the $R_L$ data
is due to small resonant contributions in the longitudinal channel.  MAID07
also provides a fairly good description of $R_T$ at \qsq=1.6 \gevsq, although
it undershoots the $R_T$ at \qsq=0.6, 1.0 \gevsq.  The overshoot at \qsq=2.45
\gevsq\ is probably due to the significant extrapolation from the optimized
parameter region $W<2$ GeV.  

We further investigated the impact of resonances in the MAID07 model on the
$\pi^-/\pi^+$ ratios.  
With all resonances turned off (Born term and meson exchange contributions on),
the model gives $R_L\approx 1$ and $R_T$ far below the data ($R_T\approx$0.5 at
\qsq=0.6, 1.0 \gevsq, $R_T\approx$0.2 at \qsq=1.6, 2.45 \gevsq).  Even though
the data are acquired near $W=2$ GeV or higher, turning on only the
P$_{33}$(1232) resonance has a significant effect on $R_T$ (increasing it to
$\approx$1.5 at \qsq=0.6, 1.0 \gevsq, and $\approx$0.8 at \qsq=1.6, 2.45
\gevsq), but it has only a small effect on $R_L$.
Progressively turning on the other resonances yields no clear trend
in the behavior of either ratio.  Curiously, turning off only the highest three
resonances, F$_{37}$(1950), P$_{31}$(1910), F$_{35}$(1905), results in virtually
no change from the nominal case.  In summary, no clear single resonance seems
to account for the global behavior of the separated ratios in the MAID07 model.
It would be extremely interesting to see the result if the model parameters
could be optimized for higher $W$.

The Goloskokov-Kroll (GK) GPD-based model \cite{gk10,gk13} is a modified
perturbative approach, incorporating the full pion electromagnetic form factor
(as determined by fits to our $F_{\pi}$ data \cite{Huber08}) in the
longitudinal channel and the $H_T$ transversity GPD dominating the transverse
channel.  The GK model is in good agreement with our $R_T$ data at
$-t_{\text{min}}$, but predicts too-flat of a $t$-dependence.  The predictions
for $R_L$ are very similar to the pion-pole dominated VGL, KM and VR models.

It is extremely important to keep in mind that the parameters in the GK model
are optimized for small skewness ($\xi<0.1$) and large $W>4$ GeV, and have not
been adjusted at all for the kinematics of our data.  This limitation becomes
apparent when comparing the GK-predicted \sigl\ and \sigt\ to our data in
Fig.~\ref{fig:xsec_models2}.  The predicted \sigt\ are too large in magnitude,
being entirely off the plotting scale at \qsq=1.0 \gevsq, and dropping very
rapidly with $-t$ to come close to the data for the highest $-t$ at \qsq=1.6,
2.45 \gevsq.  The predicted \sigl\ are generally similar to, but slightly
smaller in magnitude than the VGL, KM and VR models.  All four models use our
$^1$H $\pi^+$ data as a constraint in one manner or other.  The reasonable
agreement between the GPD-based model and our data is sufficiently encouraging
in our view to justify further effort to better describe the larger $\xi$,
smaller $W$ regime such as covered by our data.

\section{Summary}

We present $L$/$T$/$LT$/$TT$ separated cross sections for the
$^2$H$(e,e'\pi^{\pm})NN_s$ reactions, at \qsq=0.6-1.6 \gevsq, $W=1.95$ GeV and
\qsq=2.45 \gevsq, $W=2.2$ GeV.  The data were acquired with the HMS+SOS
spectrometers in Hall C of Jefferson Lab, with the exclusive production of a
single pion assured via a missing mass cut.  The separated cross sections have
typical statistical uncertainties per $t$-bin of 5-10\%.  The dominant
systematic uncertainties are due to HMS tracking at high rates ($\pi^-$), HMS
\u{C}erenkov blocking ($\pi^-$), cryotarget boiling at high current ($\pi^+$),
spectrometer acceptance modeling, radiative corrections, pion absorption and
decay.  These data represent a substantial advance over previous measurements,
which were either unseparated at \qsq=0.7 \gevsq\ \cite{Brauel1}, or separated
but over a limited kinematic range in the resonance region
\cite{Gaskell01,Gaskellthesis}.

In comparison to our previously published $\pi^+$ data from $^1$H
\cite{Blok08}, the $\pi^+$ $L$/$T$ ratios from $^2$H are higher at \qsq=0.6,
1.0 \gevsq\ but fall more steeply with $-t$, are nearly the same as from $^1$H
at \qsq=1.6 \gevsq, and lower at \qsq=2.45 \gevsq.  In contrast, the $\pi^-$
longitudinal cross sections are lower than for $\pi^+$ at \qsq=0.6, 1.0 \gevsq,
but the drop with increasing \qsq\ is less drastic and by \qsq=2.45 \gevsq\ the
$\pi^-$ $L$/$T$ ratio is slightly more favorable than for $\pi^+$.  If this
trend continues to higher \qsq, this larger $L$/$T$ ratio would benefit future
planned $L$/$T$-separations of the $^2$H$(e,e'\pi^-)pp_s$ reaction \cite{12gev}
due to a smaller error magnification factor.  \siglt\ is nearly zero for all
kinematic settings, and we also observe a significant suppression of
\sigtt\ compared to \sigt, particularly at \qsq=2.45 \gevsq.
 
Our data for $R_L$ trend toward 0.8 at low $-t$, indicating the dominance of
isovector processes in forward kinematics, which is consistent with our earlier
findings when extracting the pion form factor from $^1$H data at the same
kinematics \cite{Huber08}.  Although higher order corrections in the
longitudinal cross section are expected to be quite large even at \qsq=10
\gevsq, these corrections may largely cancel in the ratios of longitudinal
observables such as $R_L$ \cite{Belitsky,frankfurt}.  Since the transverse
target asymmetry is difficult to separate from significant non-longitudinal
contaminations at $Q^2=5-10$ \gevsq, $R_L$ may be the only practical ratio for
constraining the polarized GPDs.  In addition to the longitudinal cross
section, $R_L$ is one of the few realistically testable predictions of the GPD
model, particularly if higher order corrections cancel at a relatively low
value of \qsq\ of 2.45 \gevsq.

The evolution of $R_T$ with $-t$ shows a rapid fall off with apparently very
little \qsq-dependence above $-t=0.1$ \gevsq\ within the range covered by our
data.  Even the old photoproduction data above $-t$=0.15 \gevsq\ from DESY
\cite{heide} follow this universal curve.  The $R_T$ value at the highest $-t$
is consistent with $s$-channel quark knockout.  However, it is
unclear if this indicates a transition from nucleon and meson degrees of
freedom to quarks and gluons, as such quark-partonic behavior is at variance
with theoretical expectations of large higher twist effects in exclusive
measurements \cite{Berger} and the MAID \cite{maid07} results suggest
important soft effects.  Measurements at larger values of $-t$ and \qsq\ and
further theoretical work are clearly needed to better understand the observed
ratios.  If $R_T$ is still $\simeq$1/4
to $\pm$10\% at higher \qsq\ and similar $x_B$, the hypothesis of a quark
knockout reaction mechanism will be strengthened since there is no natural
mechanism for generating $R_T$=1/4 in a Regge model over a wide range of \qsq.
Since $R_T$ is not dominated by the pion pole term, this observable is likely
to play an important role in future transverse GPD programs planned after the
completion of the JLab 12 GeV upgrade.  The larger energy bites will also
permit simultaneous separations of electroproduction of other exclusive
transitions, such as $\gamma_v+N\rightarrow K^+\Lambda$ and $\Sigma$
\cite{12gev-K}.

\begin{acknowledgments}
The authors thank Drs. Goloskokov and Kroll for the unpublished model
calculations at the kinematics of our experiment, and Drs. Guidal, Laget, and
Vanderhaeghen for modifying their computer program for our needs.  This work is
supported by DOE and NSF (USA), NSERC (Canada), FOM (Netherlands), NATO, and
NRF (Republic of Korea).  Additional support from Jefferson Science Associates
and the University of Regina is gratefully acknowledged.  At the time these
data were taken, the Southeastern Universities Research Association (SURA)
operated the Thomas Jefferson National Accelerator Facility for the United
States Department of Energy under contract DE-AC05-84ER40150.

\end{acknowledgments}


\begin{thebibliography}{99}
\bibitem{Ber70} F. A. Berends, Phys. Rev. D {\bf 1}, 2590 (1970).
\bibitem{Gut72} F. Gutbrod and G. Kramer, Nucl. Phys. {\bf B49}, 461 (1972).
\bibitem{VGL1} M. Vanderhaeghen, M. Guidal, and J.-M. Laget, Phys. Rev. C {\bf
57}, 1454 (1998).
\bibitem{Milana} C.E. Carlson, J. Milana, Phys. Rev. Lett. {\bf 65},
1717 (1990).
\bibitem{raskin} A.S Raskin, T.W Donnelly, Annals of Physics {\bf 191}, 78
  (1989).
\bibitem{Brauel1} P. Brauel, {\em et al.}, Z. Physik C {\bf 3} 101 (1979);\\
  M. Schaedlich, Dissertation des Doktorgrades, Universitaet Hamburg (1976),
  DESY F22-76/02 November 1976.
\bibitem{collins} J.C. Collins, L. Frankfurt, M. Strikman, Phys. Rev. D {\bf
56}, 2982 (1997).
\bibitem{eides} M.I. Eides, L.L. Frankfurt, M.I. Strikman, Phys. Rev. D {\bf
59}, 114025 (1999).
\bibitem{ahmad} S. Ahmad, G.R. Goldstein, S. Liuti, Phys. Rev. D {\bf 79},
  054014 (2009).
\bibitem{gk10} S.V. Goloskokov, P. Kroll, Eur. Phys. J. C {\bf 65}, 137 (2010).
\bibitem{Belitsky} A.V. Belitsky, D. Muller. Phys. Lett. B {\bf 513}, 
349 (2001).
\bibitem{gk13} S.V. Goloskokov, P. Kroll, Eur. Phys. J. A {\bf 47}, 112 (2011);
  and Private Comminication, 2013.
\bibitem{VGL} M. Guidal, J.-M. Laget, M. Vanderhaeghen, Nucl. Phys. {\bf A
  627}, 645 (1997).
\bibitem{Blok08} H.P. Blok, {\em et al.}, Phys. Rev. C {\bf 78}, 045202 (2008).
\bibitem{kaskulov} M.M. Kaskulov, U. Mosel, Phys. Rev. C {\bf 81}, 045202 
(2010).
\bibitem{vrancx} T. Vrancx, J. Ryckebusch, Phys. Rev. C {\bf 89}, 025203
  (2014).
\bibitem{Huber14} G.M. Huber, {\em et al.}, Phys. Rev. Lett. {\bf 112}, 182501
  (2014).
\bibitem{Volmerthesis} J. Volmer, {\it The Pion Charge Form Factor via Pion
  Electroproduction on the Proton}, Ph.D. Thesis, Vrije Universiteit Amsterdam
  (2000), JLAB-PHY-00-59.
\bibitem{Tanjathesis} T. Horn, Ph D. thesis, University of Maryland (2006),
  JLAB-PHY-06-638.
\bibitem{chambers} O.K. Baker, {\em et al.}, Nucl. Instr. Meth. {\bf A367}, 92
  (1995).
\bibitem{volmer} J. Volmer, {\em et al.}, Phys. Rev. Lett. {\bf 86}, 1713 
(2001).
\bibitem{tadevosyan} V. Tadevosyan, {\em et al.}, Phys. Rev. C {\bf 75}, 
055205 (2007).
\bibitem{hornt} T. Horn, {\em et al.}, Phys. Rev. Lett. {\bf 97}, 192001
  (2006).
\bibitem{Gaskellthesis} D. Gaskell, {\it Longitudinal Electroproduction of
  Charged Pions on Hydrogen, Deuterium, and Helium-3}, Ph.D. Thesis, Oregon
  State University (2001), JLAB-PHY-01-61.
\bibitem{Koltenukthesis} D.M. Koltenuk, {\it Electroproduction of Kaons on
  Hydrogen and Deuterium}, Ph.D. Thesis, University of Pennsylvania (1999).
\bibitem{ent00} R. Ent, B.W. Filippone, N.C.R. Makins, R.G. Milner,
  T.G. O'Neill, D.A. Wasson, Phys. Rev. C {\bf 64}, 054610 (2001).
\bibitem{beb78} C.J. Bebek, {\em et al.}, Phys. Rev. D {\bf 17}, 1693 (1978).
\bibitem{frankfurt} L.L. Frankfurt, M.V. Polyakov, M. Strikman,
  M. Vanderhaeghen, Phys. Rev. Lett. {\bf 84}, 2589 (2000).
\bibitem{Huber08} G.M. Huber, {\em et al.}, Phys. Rev. C {\bf 78}, 045203
  (2008).
\bibitem{Gaskell01} D. Gaskell, et al., Phys. Rev. Lett. {\bf 87}, 202301
  (2001).
\bibitem{heide} P. Heide, {\em et al.}, Phys. Rev. Lett. {\bf 21}, 248 (1968).
\bibitem{nachtmann} O. Nachtmann, Nucl. Phys. {\bf B115}, 61 (1976).
\bibitem{Gao} L.Y. Zhu, {\em et al.}, Phys. Rev. Lett. {\bf 91}, 
022003 (2003);\ Phys. Rev. C {\bf 71}, 044603 (2005).
\bibitem{huang00} H.W. Huang, P. Kroll, Eur. Phys. J. C {\bf 17}, 423 (2000).
\bibitem{kroll02} P. Kroll, hep-ph/0207118.
\bibitem{brodsky73} S.J. Brodsky, G.R. Farrar, Phys. Rev. Lett. {\bf 31}, 1153
  (1973).
\bibitem{maid98} D.~Drechsel, O.~Hanstein, S.S.~Kamalov, L.~Tiator,
  Nucl.\ Phys.\ A {\bf 645}, 145 (1999)
\bibitem{maid07} D.~Drechsel, S.S.~Kamalov, L.~Tiator, Eur.\ Phys.\ J.\ A {\bf
  34}, 69 (2007).
\bibitem{12gev} G.M. Huber, D. Gaskell, {\em et al.}, Jefferson Lab
  Experiment E12-06-101; T. Horn, G.M. Huber, {\em et al.}, Jefferson Lab
  Experiment E12-07-105.
\bibitem{Berger} E.L. Berger, Phys. Lett. {\bf 89B}, 241 (1980).
\bibitem{12gev-K} T. Horn, G.M. Huber, P. Markowitz, {\em et al.},
  Jefferson Lab Experiment E12-09-011.
\end{thebibliography}
\end{document}